\documentclass[runningheads]{llncs} 

\usepackage{graphicx}
\usepackage[usenames,dvipsnames,svgnames,table]{xcolor}
\usepackage{mathtools}
\usepackage{amssymb,amsmath}
\usepackage{ifthen}
\usepackage{color}
\usepackage{xspace}
\usepackage{listings}
\usepackage{float}
\usepackage{makecell}
\usepackage{array}
\usepackage[utf8]{inputenc}
\usepackage{textcomp}
\usepackage{enumitem}
\usepackage{booktabs}
\usepackage[final]{pdfpages}
\usepackage{tikz}
\usepackage[htt]{hyphenat}
\usepackage[scriptsize]{subfigure}
\usepackage[small,bf]{caption}
\usepackage{threeparttable}
\newcommand*\circled[1]{\tikz[baseline=(char.base)]{
            \node[shape=circle,fill,inner sep=0.5pt] (char) {\textcolor{white}{#1}};}}

\usepackage{pifont} 
\usepackage{multirow}
\usepackage{wrapfig}
\usepackage{siunitx}
\usepackage{pslatex}

\usepackage{hyphenat}
\newboolean{showcomments}
\setboolean{showcomments}{true}
\ifthenelse{\boolean{showcomments}}
{ \newcommand{\mynote}[3]{
    \fbox{\bfseries\sffamily\scriptsize#1}
    {\small$\blacktriangleright$\textsf{\emph{\color{#3}{#2}}}$\blacktriangleleft$}}}
{ \newcommand{\mynote}[3]{}}

\newcommand{\tz}{\textsc{TrustZone}\xspace}
\newcommand{\arm}{\textsc{Arm}\xspace}
\newcommand{\optee}{\textsc{Op-Tee}\xspace}

\begin{document}
\title{On The Performance of ARM TrustZone\thanks{This is a post-peer-review, pre-copyedit version of an article published in "Distributed Applications and Interoperable Systems" (DAIS), 2019. The final authenticated version is available online at \url{https://doi.org/10.1007/978-3-030-22496-7_9}}\\
\normalsize(Practical Experience Report)}
\titlerunning{On The Performance of ARM TrustZone}
\author{Julien Amacher \and Valerio Schiavoni}
\institute{
Universit\'e de Neuch\^atel, Switzerland,
\email{first.last@unine.ch}
}
 
\maketitle
\begin{abstract}
The \tz technology, available in the vast majority of recent \arm processors, allows the execution of code inside a so-called \emph{secure world}.
It effectively provides hardware-isolated areas of the processor for sensitive data and code, \emph{i.e.}, a trusted execution environment (\emph{TEE}).
The \optee framework provides a collection of toolchain, open-source libraries and secure kernel specifically geared to develop applications for \tz.
This paper presents an in-depth performance- and energy-wise study of \tz using the \optee framework, including secure storage and the cost of switching between secure and unsecure worlds, using emulated and hardware measurements.
\end{abstract}


\section{Introduction}
\label{sec:intro}
\begin{figure}[t]
\centerline{\includegraphics[scale=0.19]{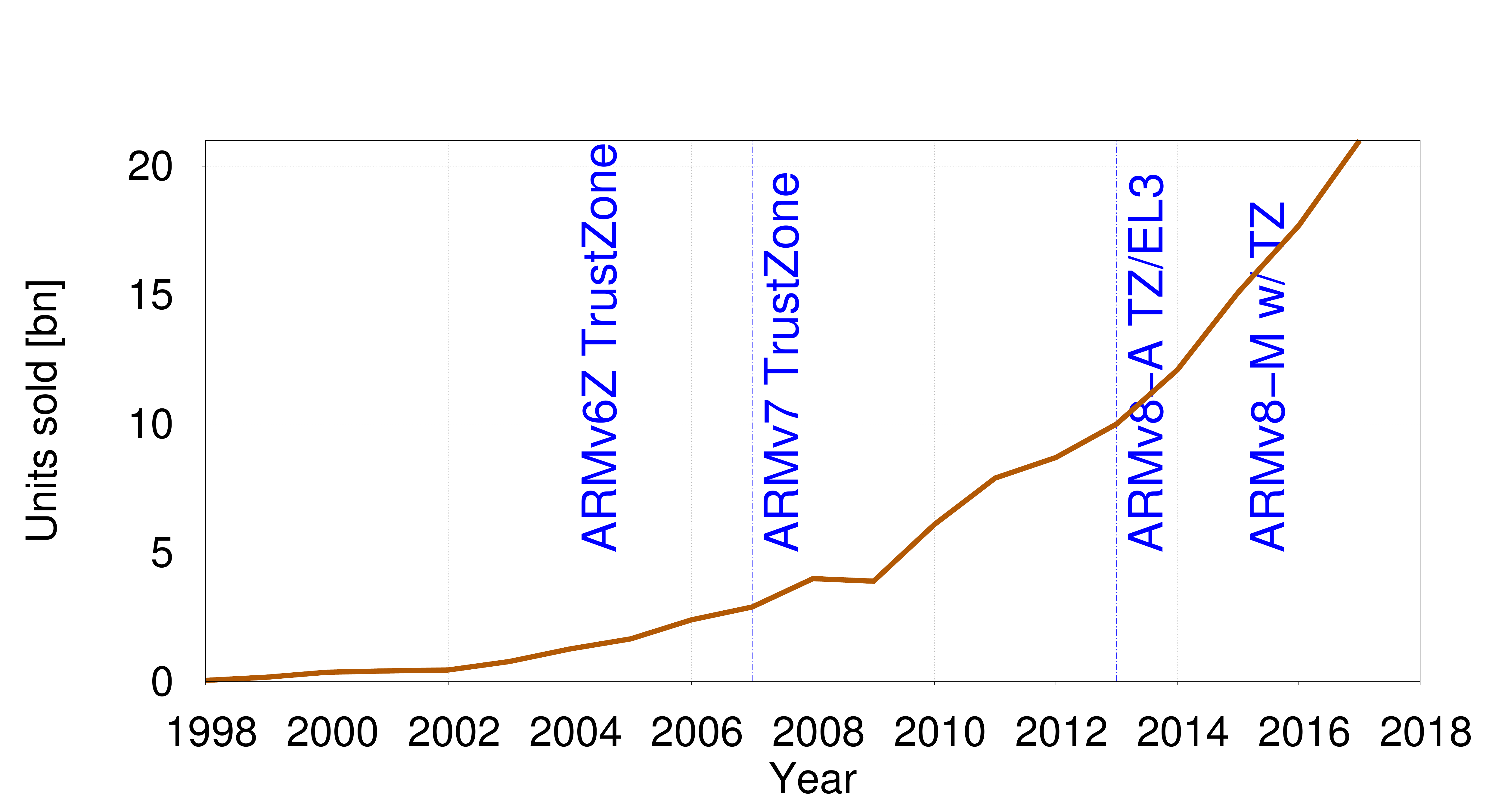}}
\caption{Sales and popularity of ARM processors in the last 20 years~\cite{armfinancianresults,armeverywhere}}\label{fig:arm_sales}
\end{figure}

Internet of Things (IoT) devices are expected to offer the pervasive computing that was promised at its advent~\cite{gartnerleadingtheiot2017}.
The economic impact of the IoT ecosystem has created many new business opportunities and is expected to continue growing rapidly.
As a result, the number of devices owned per user is anticipated to increase up to 26 by 2020~\cite{barbosa2017safethings}.
\arm, expects 275bn active devices by 2025 - a $11\times$ improvement over 2019~\cite{arm100bn} - while already having sold 100bn processors.
For instance, Figure~\ref{fig:arm_sales} reports the sales for \arm processors in the last 20 years.

These IoT devices gather, distribute and process information on their own, effectively pushing intelligence to edge devices.
Due to their nature, these devices are mostly nomad: easy to relocate, designed as wearable, embedded in vehicles or left in remote locations.
As such, assets need to be protected from attackers, in particular those easily subject to physical tampering.
Hence, ensuring that confidential data is processed in a secure manner, even in hostile environments, remains a challenging prerequisite for such devices.
Indeed, an attacker with physical access can relatively easily inspect and modify the execution workflow of any program.
Nowadays, even more disturbing attacks not requiring physical access are surfacing~\cite{lipp2018nethammer}, reinforcing the need to exploit hardware-based security mechanisms when available.
Hardware-based protections offer an additional security layer, by physically separating processing of secure and non-secure data components.
These can be dedicated processing chips (hardware security modules --HSM--), or regular chips to which security extensions were added.
Examples of the latter include Intel's \textit{Software Guard Extensions} (\emph{i.e.}, SGX~\cite{intelsgx}) since the Skylake architecture (2015), or \arm's \tz\cite{trustzone} since ARMv6 (2008).

\arm devices are often battery-powered and must therefore make optimal use of their limited energy capacity.
This is especially true nowadays, when battery capacity is becoming the limiting factor when deploying new functionalities.
Despite the availability of such devices on the market, to the best of our knowledge we could not find a public study on the performance and energy-related consumption for these security extensions.

The contributions of this work are as follows.
We begin by providing the first public experimental analysis of the performance and energy requirements of the \tz security extensions based on hands-on metrics.
Second, we report on the advantages and limitations of \optee~\cite{opteeos}, an open-source framework that supports \tz.
Third, we provide a methodology to extend the kernel of \optee in order to offer new syscalls inside \tz. 
We leverage this methodology to implement two new additional syscalls, \emph{e.g.}, to fetch thermal metrics and for secure time measurements in the \tz.
Finally, we report on our in-depth experimental analysis along several dimensions (including energy) of the current secure processing capabilities offered by some widely popular IoT devices (\emph{i.e.}, Raspberry Pi) shipping \tz processors.
Our results are put into perspective by comparing them against an emulated environment aware of the \tz extensions.

The paper is organized as follows.
\S\ref{sec:background} describes the \tz architecture and key concepts of world isolation.
\S\ref{sec:methodology} explains how the kernel was extended to expose new syscalls within \tz, how all the data was gathered, as well as the hardware and software tools that were developed.
\S\ref{sec:evaluation} presents our in-depth evaluation using real hardware and under emulation, for several hardware components (\emph{e.g.} CPU, memory, secure storage) and metrics (\emph{e.g.} performance, energy and power consumption).
We discuss some lessons learned in \S\ref{sec:lessons}, before concluding in \S\ref{sec:conclusion}.
\vspace{-15pt}
\section{Background}\label{sec:background}
This section provides some background on \tz.
First we define a few terms used throughout this paper.
\S\ref{subsec:tz} describes \tz's main mechanisms and limitations, while  \S\ref{subsec:optee} introduces \optee.

\textbf{Rich Execution Environment.} The REE (or \emph{normal world}) is the regular, non-secure operating system of a device.
The memory, registers, and caches are not isolated or protected by any hardware mechanism.
Typically, the REE is not focused on security and is difficult to review for security vulnerabilities, due to its large size and complexity.

\textbf{Trusted Execution Environments.} Also called TEE or \textit{secure OS}, it is the so-called \emph{secure world} operating system part of the \tz specifications.
It complies with the GlobalPlatform's TEE System Architecture specifications~\cite{teeinternalcorespecs}, a set of operations offered to secure applications.
These include interactions with persistent (secure) storage~\cite[Chapter~5]{teeinternalcorespecs}, memory ~\cite[Chapter~4.11]{teeinternalcorespecs}, and cryptographic operations~\cite[Chapter~6]{teeinternalcorespecs}.
As such, a secure application can easily be ported to another platform, due to the standardized nature of available services.
Similar to what a non-secure operating system offers to its running applications, the \ts TEE offers access to special services only available to secure applications (such as the secure storage feature, which we evaluate).
This environment has a small footprint, contrary to a full-fledged operating system, and only implements the very minimal set of features required to operate.
Its small size  makes it simpler to review for security vulnerabilities, as any could potentially compromise all secure applications. 

\textbf{Trusted Application.} A trusted application (\textit{TA}), also called secure application is designed to be run exclusively inside the secure world.
It uses services provided by the TEE kernel to access resources, specifically: (1)  disk via the secure storage subsystem exclusively,  (2) TCP/IP sockets, (3) memory allocation, (4) other custom services.
Trusted applications provide services to either standard userland programs or other TAs. \optee expects TAs to be written in C.

\subsection{\tz in a nutshell}\label{subsec:tz}
\begin{figure}[!t]
\centerline{\includegraphics[scale=0.43,trim=0 0cm 0cm 0cm]{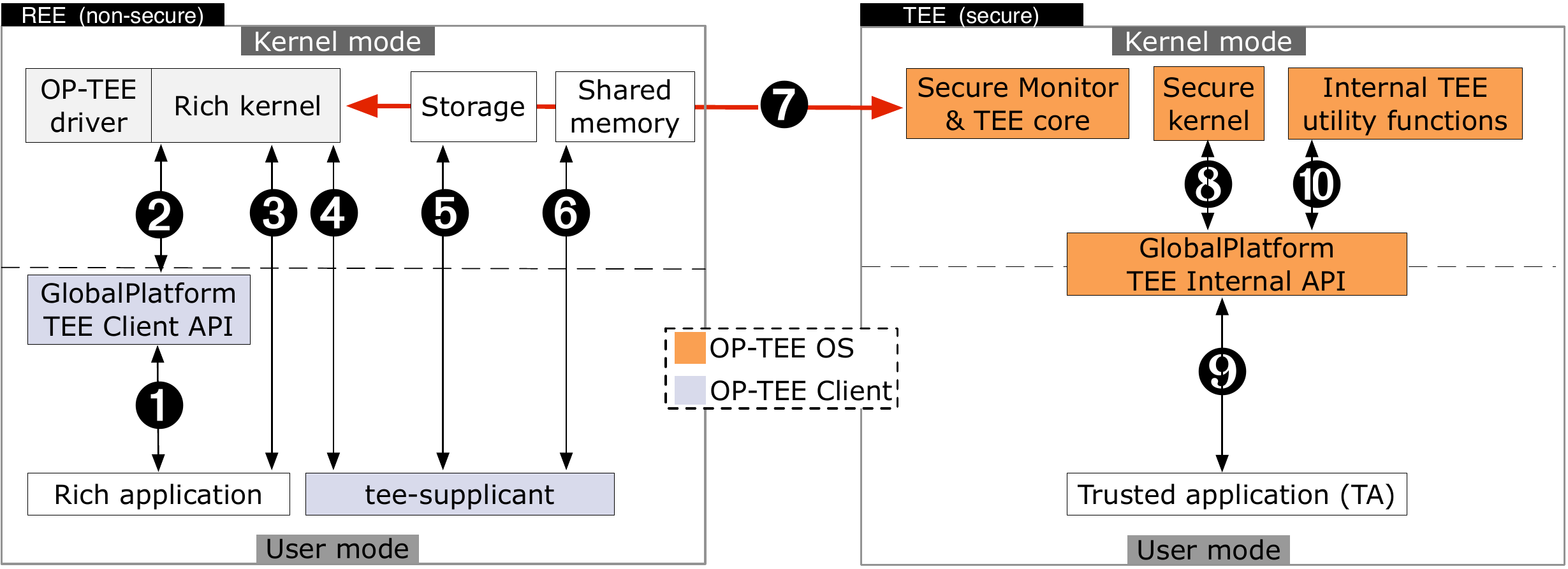}}
\caption{\tz components and interaction workflow.}\label{fig:trustzone_components}
\end{figure}

This section describes the main components of the \tz architecture, also depicted in Figure~\ref{fig:trustzone_components} alongside their interfaces.

\textbf{Overview.} \tz is a hardware feature implemented in recent \arm processors.
It enables physical separation of different execution environments, namely TEE and REE.
Its working principle is very similar to a hypervisor, the main difference being that no emulation is performed and that all isolation is offered at the hardware level.
Both secure (TEE) and normal worlds (REE) share the underlying physical processor. 
The secure world has unrestricted access to memory regions, hardware and devices. 
This is realized by using an additional addressing line, the \texttt{NS} (Non Secure) bit.
Hardware checks performed by the \textsc{Tzasc} (\tz Address Space Controller)~\cite{armtzc,secloakfull} determines, if the access is authorized based on this NS-bit. 

\textbf{Memory.} Parts of the memory can be isolated for exclusive use by the secure world by means of special hardware support.
The memory management unit (MMU) is secure-world aware, and secure and non-secure descriptors are stored alongside each other.
The differentiation is done by the \emph{Non-secure TLB ID}~(NSTID)~\cite{nstid}, an extra bit of the TLB.
The secure applications (TAs) must fit in the on-chip memory.
Due to high costs of the secure memory, it is usually limited in size, in the order of 3-5MB. 
Hence, TAs are expected to have small memory footprints and only contain the minimal subset of features required.
Clearly, this reduces the attack surface exposed by TAs. 

\textbf{Interrupts.} The \emph{Fast Interrupt} (FIQ) secure interrupt mode is used exclusively by devices residing in a memory region allocated to the secure world.
As such, regular interrupts (IRQ), which are of lower priority, cannot be used to prevent the secure world from executing, in particular if a physical secure clock (\emph{i.e.}, RTC) is used.
Secure clocks are crucial to ensure a TA is safely executed: an external clock is a common attack vector and can be easily tampered with~\cite{clockglitchncc2015}. Latest \arm processors include secure  clocks.

\textbf{World Switching.} Switching between worlds requires the state of the processor to be saved and then restored, respectively when entering and exiting a new world.
Processor registers are saved by the monitor when entering, and restored when leaving the secure world.
The NS-bit is changed accordingly.
Normal world applications use \tz indirectly, by invoking functionalities implemented in a dedicated TA.
When in PL-1~\cite{smcallingconvention,armsystemcallsel1} privilege level, a special hardware instruction, \emph{Secure Monitor Call} (SMC), allows switching between worlds. 
Recent Cortex-A processors~\cite{greenhalgh2011big} support SMC calls by the kernel in the normal world.
Entry to a different world (from secure to unsecure and vice versa) is done on a core-basis, thus limiting the parallel execution of TAs to the number of available cores.
To enter the secure world, a kernel thread executes the monitor, which in turn issues the SMC instruction to the CPU~\cite{armsmc,genericentrysmc}.
Calls to SMC by a processor not in kernel mode trigger an undefined exception trap.
TAs can be called from userland programs residing in the REE or from other TAs. 
The latter is particularly useful to reduce code duplication and to keep the TA's attack surface minimal.
Data is passed back and forth between worlds by memory pointers or direct copies. 

\textbf{Secure storage.} \tz supports persistent data storage for TAs using secure storage.
Objects are stored encrypted on disk, and are signed for anti-tampering countermeasure.
TAs access the files in cleartext: the TEE layer runs the cryptographic stack transparently.
These files have a unique numeric name based on a counter. 
An encrypted index of files is maintained alongside the files.
Operations on the index are atomic, ensuring integrity protection by means of a hash tree data structure that guards the index.
To protect against storage replay attacks, an eMMC storage device (\emph{embedded MultiMediaCard}, a type of non-volatile, non-removable solid-state storage device~\cite{emmcexpl}) is required. 
This security feature is entirely implemented in the eMMC storage in the form of \emph{Replay Protected Memory Block} (RPMB)~\cite{7411305}.

\textbf{Key Management.} The key manager starts with a device-specific key, the \emph{Secure Storage Key} (SSK).
It is derived from two pieces of information unique to each device's processor: the chip identifier and the hardware key.
The \emph{TA Storage Key} (TSK) is a per-TA key, derived from the SSK and the TA's UUID identifier.
The \emph{File Encryption Key} (FEK) is a per-file key generated upon file creation. 
It is used to protect the file contents, including its metadata, and is encrypted using the TSK.



\textbf{Resilience to attacks.} It is of paramount importance to ensure that only trustworthy applications are deployed to the secure world.
Vulnerabilities in any TA, the TEE or a compromised secure kernel do compromise the security of the secure world.
Prevention against buffer overflow attacks in the secure world are currently only provided using basic stack canaries~\cite{opteecanaries}. 
Future support for ASLR (Address Space Layout Randomization) will improve resilience against those attacks.
Finally, there exist mitigations against Meltdown and Spectre speculative execution attacks~\cite{cve20175754,cve20175715,cve20175753,cve20183639}.
Covert data channels~\cite{teecovertchannelsprimecount} can also be used when required.

\begin {table}[t]
\centering
\begin{threeparttable}
\rowcolors{1}{gray!10}{gray!5}
\begin{tabular}{  m{3cm}  m{3cm}  m{3cm}  }
\hline
\rowcolor{gray!25}
\multicolumn{1}{c}{\textbf{Framework}} & \multicolumn{1}{c}{\textbf{License}} 				& \multicolumn{1}{c}{\textbf{Technology}} 	\\ \hline
\optee~\cite{opteeos} 						& BSD 					& \tz 	\\
Trustonic TEE~\cite{trustsonic} 	    	& Commercial 			& \tz 	\\
Open TEE~\cite{mcgillion2015open}  			& Apache License 2.0  	& \tz 	\\
OpenEnclaves~\cite{openenclave}		    	& MIT  			        & SGX1 \& \tz \\ 
TLK~\cite{tlk}  							& BSD  					& \textsc{Nvidia} Tegra \\
Android Trusty TEE~\cite{androidtrustytee}	& Apache License 2.0  	& $\tz^{\text{1}}$ \\ \hline
\end{tabular}
\begin{tablenotes}
\small
\item ${}^{\text{1}}$: emulated under Intel's VT
\end{tablenotes}
\end{threeparttable}
\caption{Existing frameworks for TEE-based applications.}\label{table:common_tee}
\end{table}

\subsection{The \optee Trusted OS}\label{subsec:optee}
While there are few options (Table~\ref{table:common_tee}) to develop applications for TEEs, we rely on \optee, 
due to its fast development cycle and native support for the \tz.

\optee is a security framework  
that includes several components: a minimal secure-world operating system (the \optee\textsc{Os}~\cite{opteeos}); the  \textit{tee-supplicant}~\cite{opteeteesupplicant}, offering normal world services to the secure world; a complete build toolchain~\cite{opteebuild}, the testing tool~\cite{opteesanitytestsuite} (\textit{OPTEE sanity testsuite}), a secure privileged layer enabling world switching, a basic REE image, and several utility functions for developers to implement TAs.
\optee is flexible and can be deployed to platforms for which there exists a manifest, that lists the dependencies required to build for the platform it describes, as well as its hardware characteristics.
Additionally, the Qemu open source emulator~\cite{qemu} allows to deploy and evaluate \optee in emulated mode on ubiquitous machines.
The TEE interface implemented in \optee is compliant with the GlobalPlatform's specifications. 

\textbf{Details.} \optee imposes a specific interface regarding TA interactions initiated from the REE.
First, a request to load the desired TA is made by passing its UUID to \emph{TEEC\_InitializeContext} which returns a context object. The UUID is defined at compile-time and must be unique amongst all TAs.
Next, this context is passed to \emph{TEEC\_OpenSession} which returns a session.
This session is then used to invoke actual services in the TA using the \emph{TEEC\_InvokeCommand}, which takes as parameters the service identifier as well as any optional parameters.
A single session can be used to call \emph{TEEC\_InvokeCommand} any number of times.
Sessions are finally closed using \emph{TEEC\_CloseSession} and ultimately, the context is closed by calling \emph{TEEC\_FinalizeContext}.
To support multiple sessions, the TA must be compiled with the \emph{TA\_FLAG\_MULTI\_SESSION} flag set.
\optee signs TAs with a private RSA key, but the toolchain does not allow a unique key per-TA (all TAs are signed with the same device key). 
Upon TA loading, the \optee core checks the integrity of the TA by verifying its signature based on its signed header.
The framework includes a minimal OS that offers services to TAs, and leverages the tee-supplicant application to access resources residing in user land.

\section{Methodology}\label{sec:methodology}
This section describes the tools and techniques used to carry out our evaluation. 
We focus on four metrics : (1) execution time for various types of benchmarks (CPU-bound, volatile and non-volatile memory), (2) power consumption under different CPU governors, (3) energy consumption, and (4) thermal behaviour of the CPU.


\begin{figure}[!t]
\centerline{\includegraphics[scale=0.5,trim=0 14.4cm 4cm 0cm]{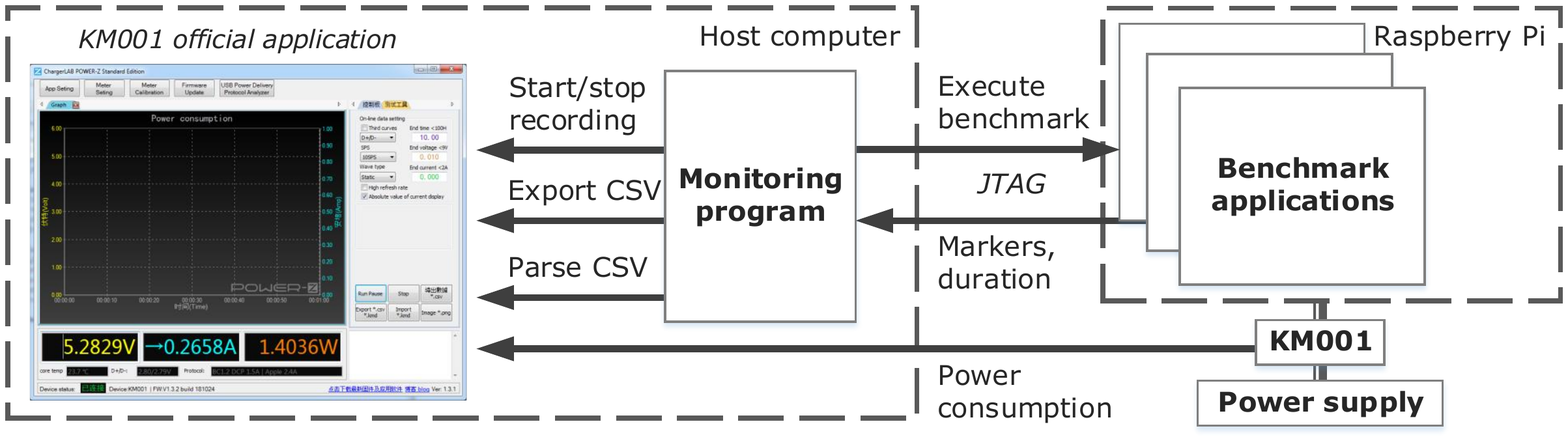}}
\caption{Experimental setup and approach used to run our measurements}\label{fig:setup}
\end{figure}	

\begin{figure}[t]
\begin{minipage}[c]{0.4\linewidth}
\includegraphics[scale=0.5,trim=0 9.5cm 15cm 0cm]{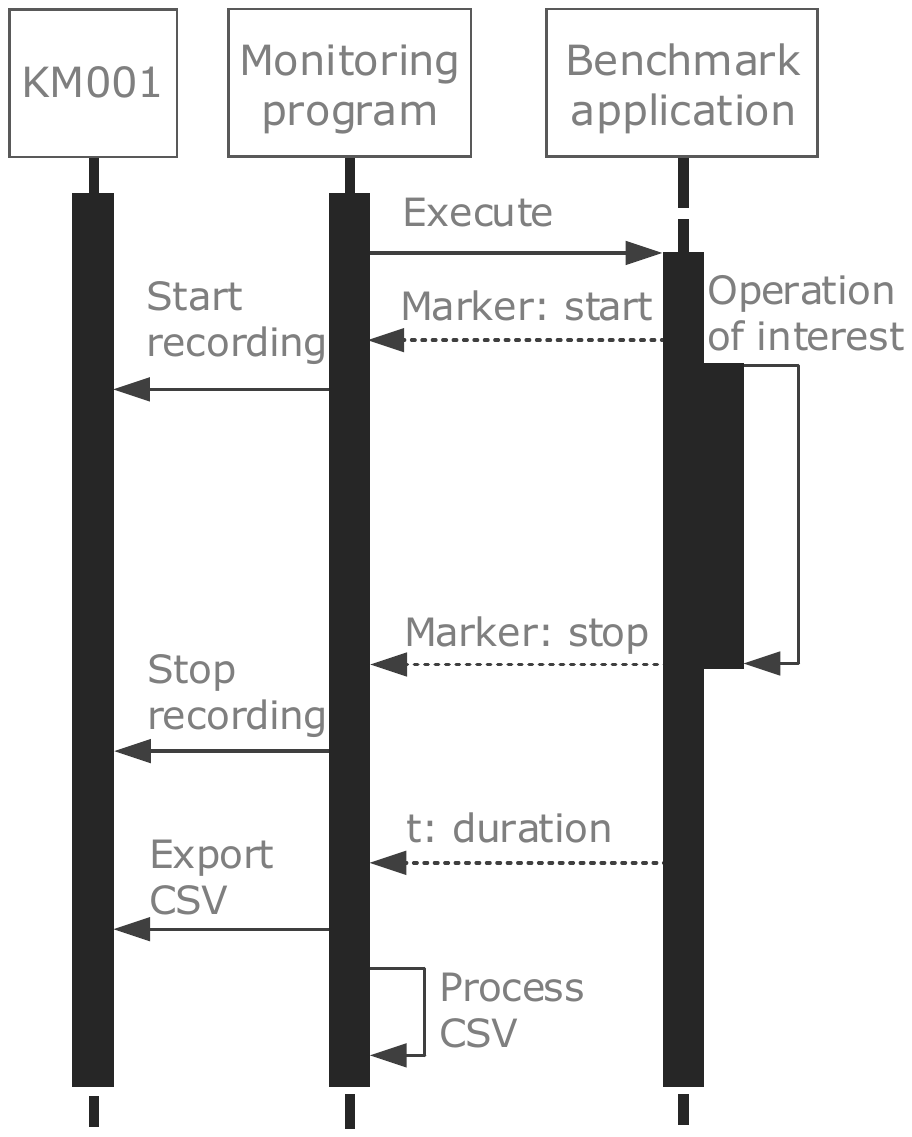}
\caption{Use of markers}
\label{fig:benchprotocol1}
\end{minipage}
\hfill
\begin{minipage}[c]{0.6\linewidth}
\includegraphics[scale=0.5,trim=0 9.5cm 11cm 0cm]{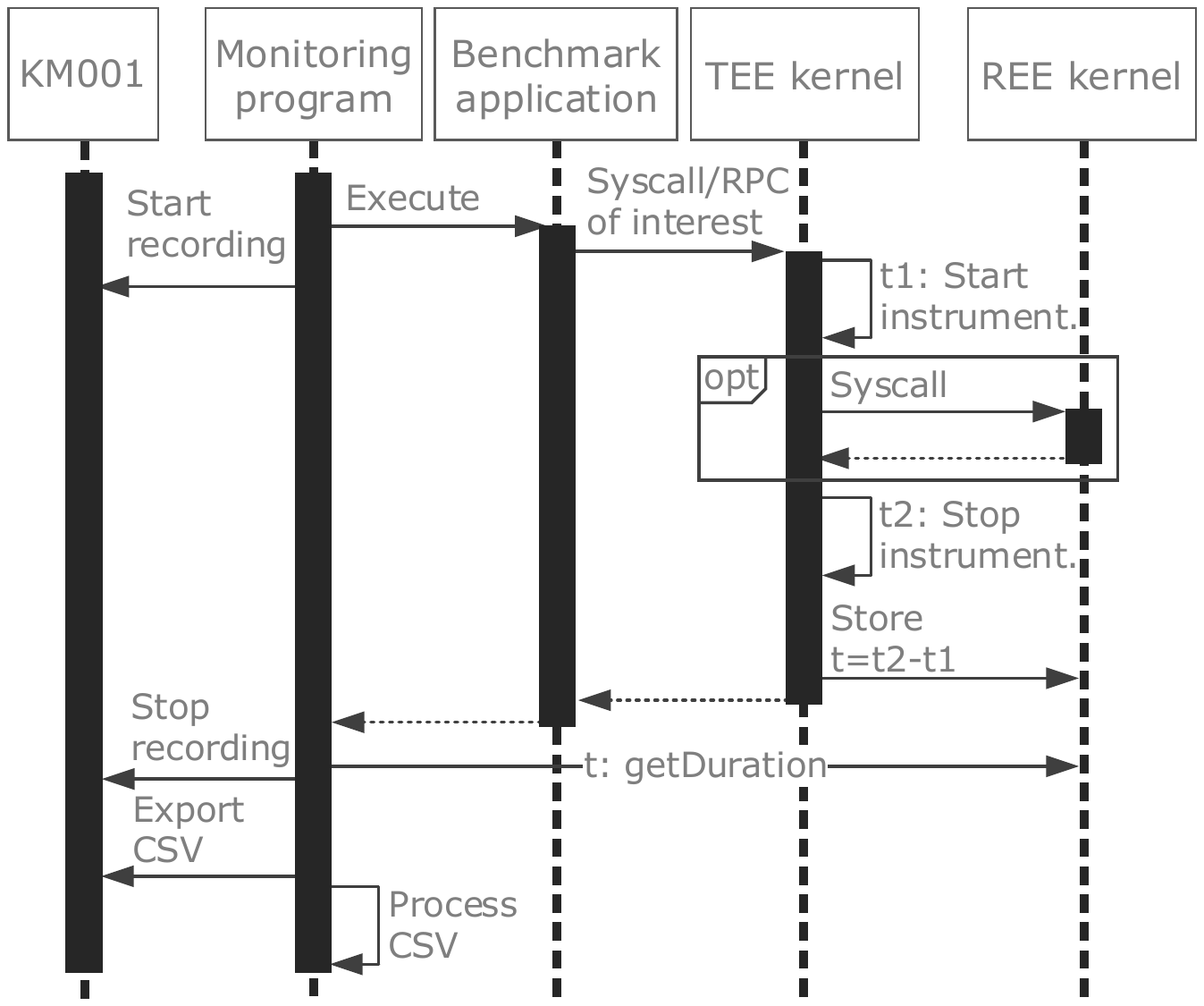}
\caption{Microbenchmarking: workflow}
\label{fig:benchprotocol2}
\end{minipage}%
\end{figure}

\textbf{Hardware Measurement Tools.}\label{subsec:tool} Energy and power measurements are carried out using a Power-Z KM001 unit~\cite{powerzkm001}, plugged in-between the USB power supply and the Raspberry Pi device.
The variant used in our testbed features two main USB ports (to provide power and one from where the power is drawn) of the current mainstream USB types (type A, micro and type C). 
In our configuration, type A is used for both input and output of power delivery. 
An additional (micro) USB port is used to 
fetch power consumption measurements.
The KM001 unit supports different USB protocols, including USB PD (Power Delivery) 2.0 and Qualcomm QC (QuickCharge) from version 2.0 up to 4.0.
This configuration allows the power used by the Raspberry Pi to be measured directly as the losses of the power supply itself are not taken into account.
We use this device to measure only power [W] and energy [Wh], for which it produces 1 record per second.
%
Unfortunately, the software (Figure~\ref{fig:setup}, left) provided by the unit manufacturer is a closed-source 32-bit Windows binary,  and the protocol used to exchange messages over USB is undocumented.
To overcome these limitations, we used the following approach. 
Specific markers (\emph{e.g.} \emph{start recording} and \emph{stop recording}) are generated during execution of benchmark applications, allowing for precise recording of areas of interest (Figure~\ref{fig:benchprotocol1}).
These markers are monitored by a custom program (on a separate node) that pilot the Windows binary (Figure~\ref{fig:benchprotocol2}).
The pilot sends automated messages to the binary instance using the Win32 API through P/Invoke (Platform Invokation Service)~\cite{microsoftinterop} issued by a monitoring program implemented in C\#.

\textbf{CPU Governors.} \label{subsec:govs} The Linux kernel supports several CPU governors~\cite{kernelgovernors}, used to adjust the frequency of each core depending on its load and temperature.
Several options exist: \texttt{powersave} and \texttt{performance} for minimum and maximum operating frequency; \texttt{ondemand} toggles between the previous two, and a more \texttt{conservative} mode that operates less aggressively; \texttt{userspace}, to manually set the CPU frequency; and \texttt{schedutil}, where the frequency is set by the scheduler.
The core frequency is increased during the execution of stressful workloads and reduced right after, for instance when  the maximum temperature is reached in order to prevent overheating.
This is different from a hardware thermal throttling, which tries to prevent damage caused by excessive heat.
The \optee kernel uses \texttt{powersave} governor by default.
This reduces heat output by reducing the frequency of the core clocks, allowing passive cooling - even without heatsink - but also negatively impacts performance.
In a compute-intensive datacenter, one would typically use the \texttt{performance} governor.
Instead, if energy constraints are important, the \texttt{powersave} mode is best suited. 
Our benchmarks consider both governors and compare them for REE and TEE executions. 

\textbf{Timing issues.}\label{subsec:benchmarks} Initially, we planned on porting \textsc{Stress-NG}~\cite{stressng} to run inside \tz.
Unfortunately this proved to be not straightforward, given its reliance on system calls not available inside the TEE kernel.
As such, we decided to implement custom ad-hoc benchmark applications.
Execution time is measured using either the \texttt{gettimeofday(2)}~\cite{gettimeofday2} or the \texttt{clock\_gettime(3)}~\cite{clockgettime3} syscall, which support the following parameters:
\begin{enumerate}
\item \texttt{CLOCK\_REALTIME}: the realtime clock of the system, can be adjusted by NTP and thus can go forward and backwards.
\item \texttt{CLOCK\_MONOTONIC}: a monotonic time since an unspecified starting point (usually system startup, as is the case with our setup)
\item \texttt{CLOCK\_PROCESS\_CPUTIME\_ID}: per-process timer
\item \texttt{CLOCK\_THREAD\_CPUTIME\_ID}: thread-specific CPU-time clock
\end{enumerate}
For our experiments we exclusively use \texttt{CLOCK\_MONOTONIC}.
Our benchmarks include the instrumentation delay, \emph{e.g.}, the overhead introduced by the measurement itself.
This is especially important from the TEE perspective (\emph{i.e.}, inside a TA) where one syscall can lead to a second one if REE needs to be accessed (\emph{e.g.}, Figure~\ref{fig:trustzone_components}-\ding{210} and Figure~\ref{fig:trustzone_components}--\ding{208}).

\textbf{Kernel and \optee modifications.}\label{subsec:kernelexts} To access and store the monotonic time and temperature from within a TA using the secure kernel, and to retrieve it later on within the REE, we extended the kernel with four new system calls: \texttt{TEE\_GetCpuTemperature, sys\_ktraceadd, sys\_ktraceget} and \texttt{sys\_ktracereset}. 

To gather the temperature measurements, we used two methods: (1) software, via thermal APIs\footnote{\texttt{/sys/class/thermal/thermal\_zone[0-9]+/temp}} and (2) external hardware sensor.
Originally, we planned on using a script to record the temperature at fixed intervals during the CPU stress tests executed by userland threads.
However, since kernel threads executing the TAs have a higher priority, the userland threads were starved and thus did not produce enough data points.
This is a typical scenario of normal world starvation occurring when TAs monopolize all cores.
We overcome this problem by accessing the CPU temperature from inside the TA, and sending it periodically to the monitoring software for safekeeping.
To use the temperature gathering syscall from within the TA, we additionally had to implement the corresponding TEE kernel syscall wrapper.
An extensive walkthrough on this process is given in Appendix~\ref{appendix:extending-kernel}.
\section{Evaluation}\label{sec:evaluation}
This section presents our in-depth evaluation and performance analysis, the main contribution of this work. 
Energy results are always presented by systematically excluding idle energy consumption, \emph{e.g.}, we only show the energy cost of the given operation.
Energy requirements are shown on a per-operation fashion. 
To prevent thermal throttling, all tests run while the onboard chip is actively cooled.

\begin{figure}[t]
\begin{minipage}[c]{0.60\linewidth}
\center
\includegraphics[scale=0.14]{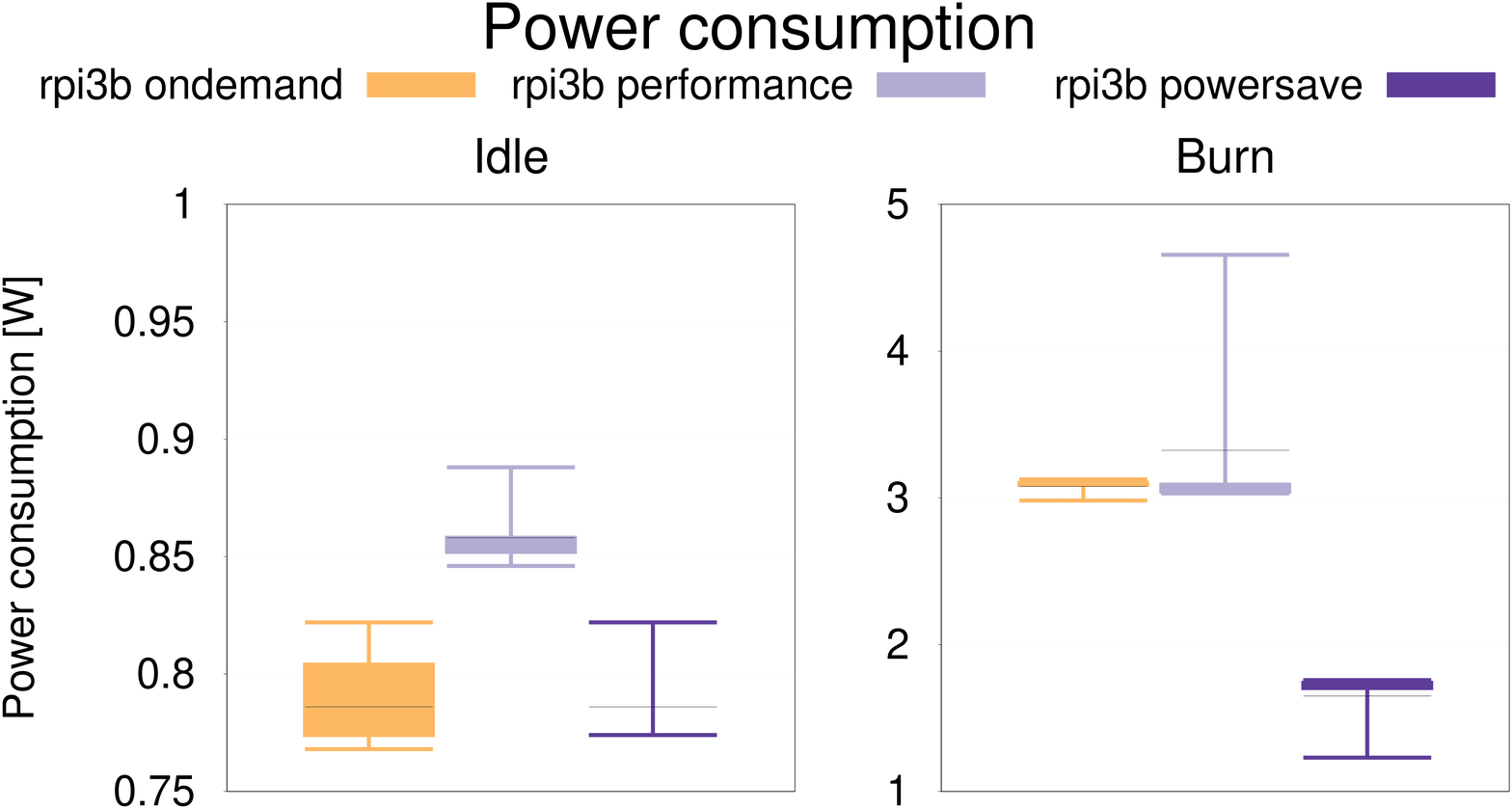}
\caption{Idle (left) and burn (right) power consumption.}\label{fig:power_consumption}
\end{minipage}
\hfill
\begin{minipage}[c]{0.38\linewidth}
\begin{table}[H]
\small
\begin{center}
\bgroup
\begin{tabular}{  m{1.55cm} m{0.5cm} m{0.75cm} m{0.5cm} m{0.75cm} }
\hline
\rowcolor{gray!25}
 & \multicolumn{2}{c}{\textbf{Idle}} & \multicolumn{2}{c}{\textbf{Burn}}\tabularnewline
\rowcolor{gray!25}
\multirow{-2}{*}{\textbf{Governor}} & \centering{\scriptsize W} & \centering{\scriptsize BTU/h} & \centering{\scriptsize W} & \centering{\scriptsize BTU/h}\tabularnewline
\hline
ondemand       & \centering{0.78} & \centering{2.66} & \centering{3.08} & \centering{10.51}\tabularnewline
performance & \centering{0.86} & \centering{2.93} &    \centering{3.32} & \centering{11.33}\tabularnewline
powersave      & \centering{0.78} & \centering{2.66} & \centering{1.65} & \centering{5.63}\tabularnewline \hline
\end{tabular}
\egroup
\vspace{0.5cm}
\caption{Average power consumptions for idle and burn experiments (see Figure~\ref{fig:power_consumption})}\label{table:power_consumption}
\end{center}
\end{table}
\end{minipage}
\vspace{-18pt}
\end{figure}

\textbf{Evaluation Settings.} We use the Raspberry Pi 3B, a popular yet representative single-board device, equipped with Broadcom BCM2837 \emph{System-On-Chip} (1GB of RAM, ARM Cortex A53 quad core running at 1.2GHz). 
For some of our measurements, we compared the hardware experiments against a modified version of the Qemu emulator provided by \optee with support for \tz~\cite{qemutzgit}.
This mimics the scenario of an Infrastructure-as-a-Service provider offering access to \arm nodes (as virtual machines) to cloud tenants without having the corresponding hardware infrastructure and thus relying on \tz virtualization~\cite{hua2017vtz}.
Qemu uses the Cortex A53 emulation profile on an Ubuntu host residing on a VMWare ESXi~\cite{vmwareesxi} machine equipped with an i7 6820HQ running at 2.7GHz.
Note that the Raspberry Pi 3B lacks support for secure boot and hardware separation of memory and peripherals~\cite{unsecurerpi3b}, hence these aspects of the \tz ecosystem could not be evaluated and are left for future work.
Finally, we do not override the default secure storage key (SSK) provided by \optee. 

\textbf{Power consumption.}\label{subsec:power} We start by measuring the idle and under-stress (\emph{burn}) power consumption of our hardware unit.
We evaluate how the three different CPU governors (\texttt{ondemand}, \texttt{performance}, and \texttt{powersave}) behave.
The idle measurements use the standard REE kernel image provided by \optee, without any user-intensive applications nor TAs running. 
Burn measurements run the prime benchmark, a single-threaded TA which computes the first 20000 prime numbers before exiting.
We run 8 instances in parallel, ensuring maximum heat output on the 4 cores. 
Measurements start 60 seconds after the benchmark instances.
Figure~\ref{fig:power_consumption} shows our results, respectively for idle (left) and burn (right) experiments. 
Table~\ref{table:power_consumption} shows the average W and BTU/h.
We use a box-and-whiskers plot: the first and third quartile are shown as a colored box, the median as horizontal black bar. Min/max values are also included. 
Results for \textit{ondemand} and \textit{powersave} are on par with the \textit{ondemand} governor, in particular when the CPU frequency is set at 600MHz.
As expected, we observe higher power consumption using the \texttt{performance} governor even in idle, as the cores are boosted up to 1.2GHz.
Overall, the board's power consumption is very low, in particular below 1W in idle mode.
%

\textbf{Load \& unload TAs.} Next, we measure the time required to load and unload a TA inside the \tz, respectively executing \textit{TEEC\_InitializeContext}~\cite[Chapter 4.5.2]{globalplatformclientapi} and \textit{TEEC\_FinalizeContext}~\cite[Chapter 4.5.3]{globalplatformclientapi} functions.
We compare results obtained with a TA of size smaller and another one of size larger than the 512kB L2 cache of the Broadcom BCM2837 processor, respectively 102kB and 517kB.
Our experiments show no significant difference between TAs of different sizes. 
\begin{wrapfigure}{r}{.6\textwidth}
\vspace{-23pt}
\centerline{\includegraphics[scale=0.15]{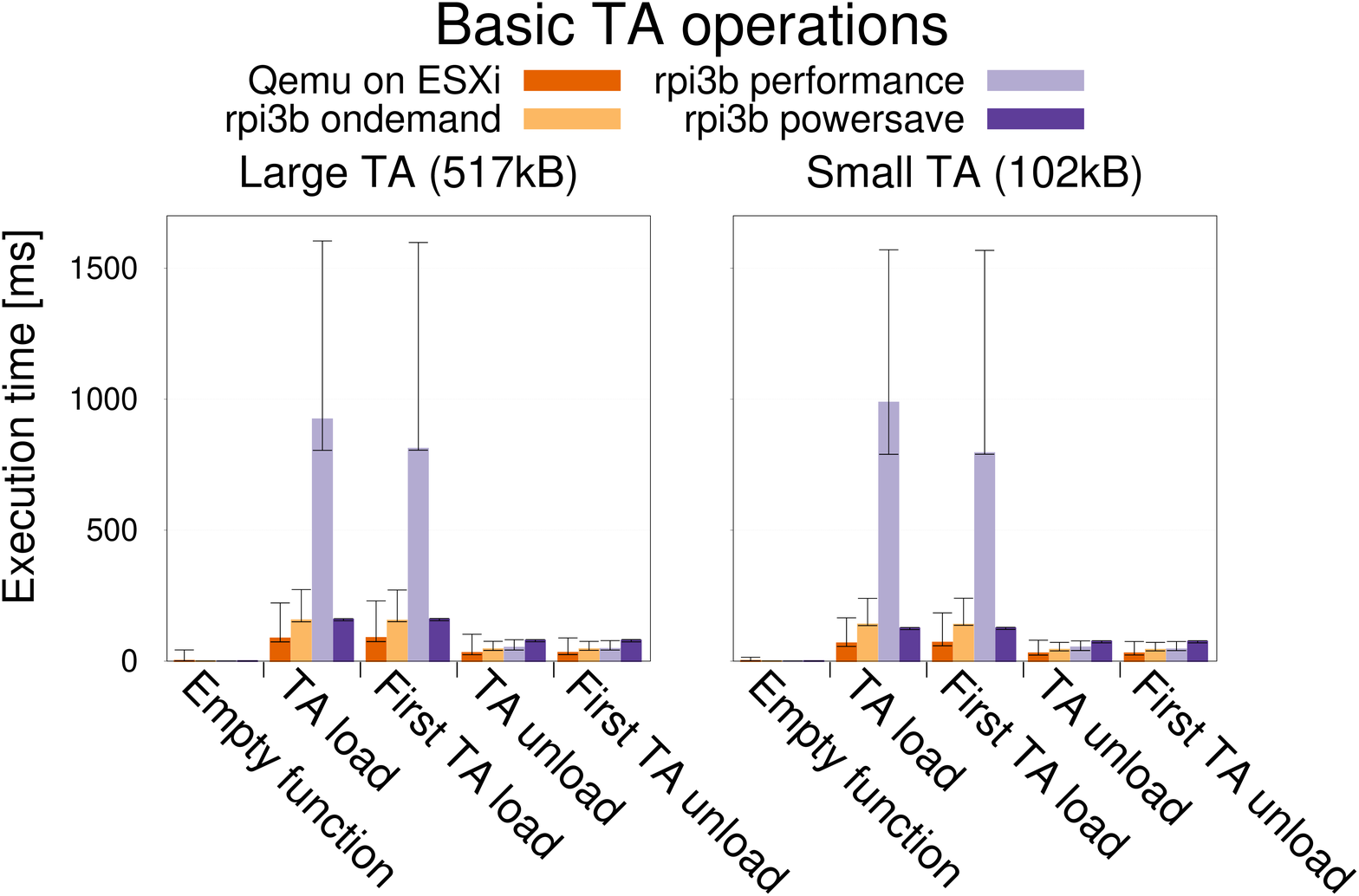}}
\caption{Basic TA operations: loading, unloading and successive calls to load/unload the same TA.}\label{fig:performance_ta_loading}
\vspace{-15pt}
\end{wrapfigure}
For each configuration, Figure~\ref{fig:performance_ta_loading} shows average and standard deviation over 10k executions.
We include the time spent to execute an empty function inside the TA once it is loaded (1.31ms), to give a baseline of comparison.

%
Surprisingly, our results do not show a significant differences on subsequent loadings compared to the first loading, despite the tee-supplicant is supposed to cache the TA code. 
We will investigate this aspect in future work.

\begin{figure}[t]
\center
\includegraphics[scale=0.15]{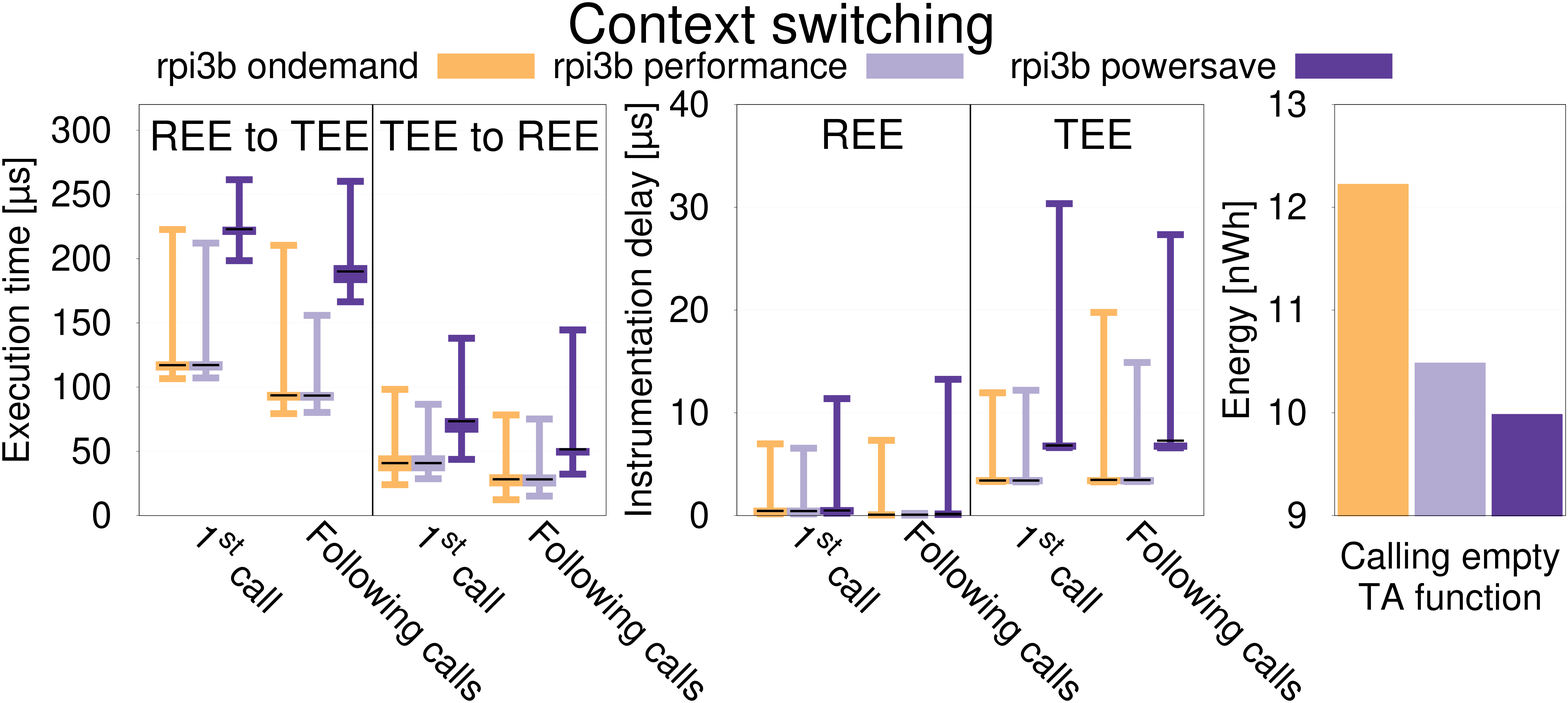}
\caption{World switching performance and energy requirements}\label{fig:performance_context_switch_and_energy_requirements}
\end{figure}
\textbf{Context (World) Switching.} Switching between worlds is a key operation when deploying applications that execute inside and outside the \tz.
To measure the switching time, we implemented an ad-hoc benchmark made by a host application and a TA.
Both programs record the monotonic time when entering and exiting the world in which they reside. 
The host issues a call to an almost empty function, which only contain time-measuring code.
Two calls are made to the TA per session, recording the time taken to switch between TEE and REE, and vice versa.
Figure~\ref{fig:performance_context_switch_and_energy_requirements} (left) shows these results.
To evaluate possible caching effects, we also include the results obtained for all the calls following the first one.
As expected, it is more time-consuming to switch from the REE to the TEE (110\si\micro s with the performance-oriented governors) than the opposite (47\si\micro s).
The instrumentation delay (Figure~\ref{fig:performance_context_switch_and_energy_requirements}, center) is the difference between two consecutive calls to the time measurement function.
An increased instrumentation delay is observed in the TEE compared to the REE, due to the additional world switch.
Finally, we also evaluate the energy spent for calling an empty TA function from the REE (Figure~\ref{fig:performance_context_switch_and_energy_requirements}, right).
The timer starts and stops when leaving and re-entering the REE, respectively.
The \textit{ondemand} governor is the most energy-eager (up to 12.1 nWh), while \textit{powersave} is the most energy efficient.





\begin{figure}[t]
\begin{minipage}[c]{0.6\linewidth}
\center
\includegraphics[scale=0.17]{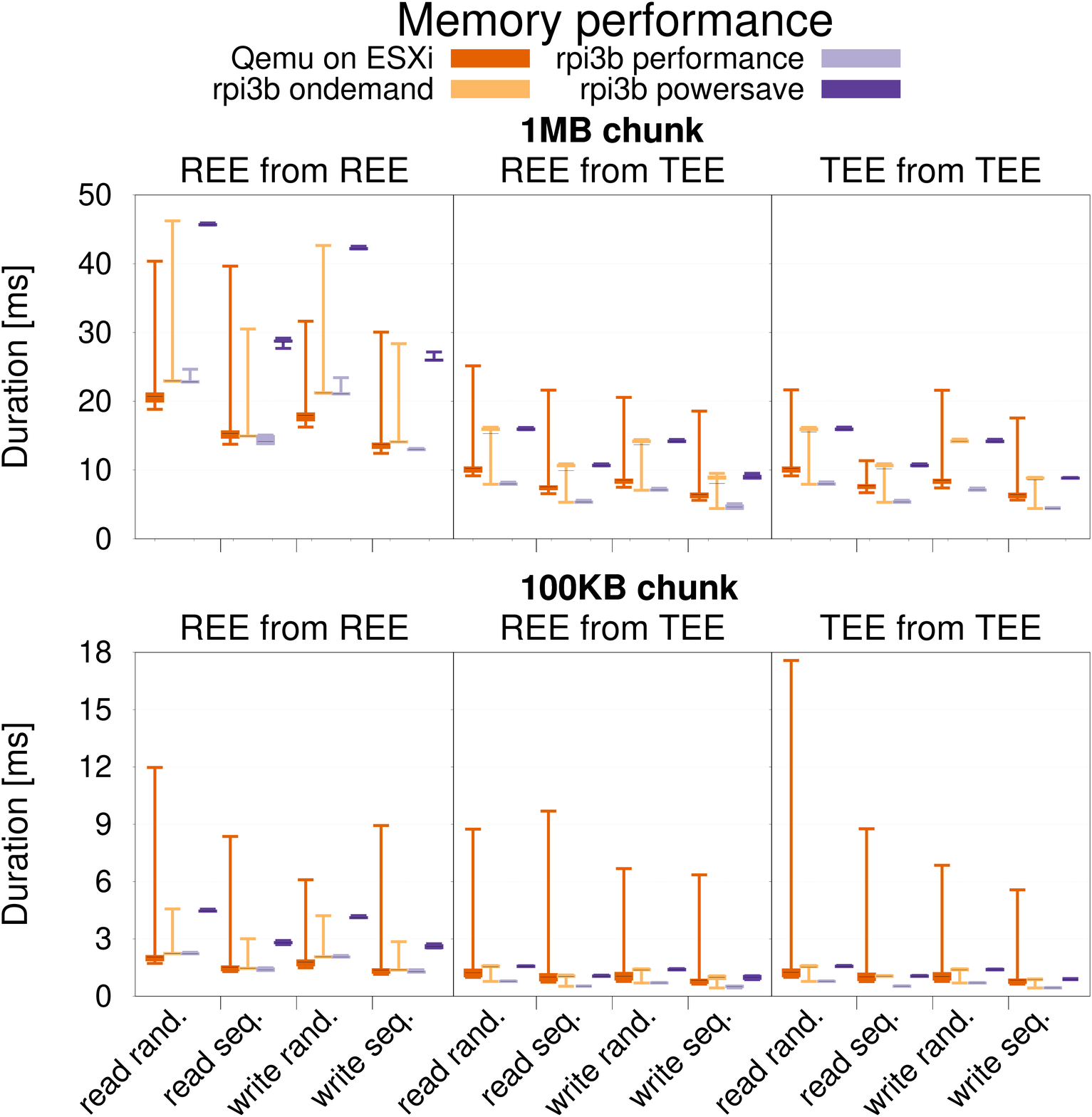}
\caption{Benchmark for memory ops}\label{fig:performance_mem_1MB_100KB}
\end{minipage}
\hfill
\begin{minipage}[c]{0.38\linewidth}
\center
\includegraphics[scale=0.17]{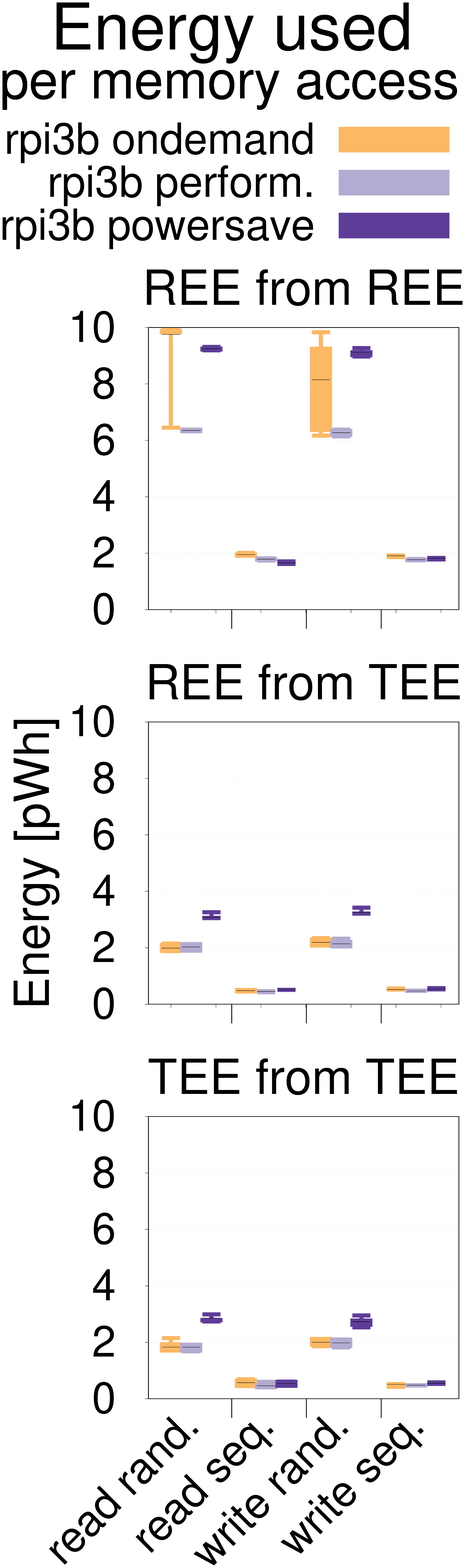}
\caption{Energy: memory accesses}\label{fig:power_mem}
\end{minipage}
\end{figure}


\textbf{Volatile Memory.} Next, we consider simple in-memory operations (\emph{e.g.}, read and write, sequential or at random), for two different sizes of volatile memory (1MB and 100KB) used by the REE and the TEE.
We consider inter- (REE$\leftarrow$TEE) and intra-world (\emph{e.g.}, REE$\leftrightarrow$REE, TEE$\leftrightarrow$TEE) memory readings, as \tz restrictions prevents reading TEE memory from the REE.
We compute the average and standard deviation over 100 run, always using the high-resolution monotonic counter.
Figure~\ref{fig:performance_mem_1MB_100KB} shows our results, for the Raspberry Pi device with 3 CPU governors and using Qemu.
Performance of accessing a single byte in TEE memory from the TEE is on par with accessing REE memory from the TEE, on average 0.01\si\micro s, around $2\times$ under emulation.
Interestingly, using memory from within the TEE is also less energy eager (Figure~\ref{fig:power_mem}), also verified by the cost of the single operations in the various configurations. 
We observe how the operations in the TEE$\leftrightarrow$TEE case are on average $2\times$ faster on bare metal and $1.2\times$ under emulation than in the other cases.


\begin{figure}[t]
\centerline{\includegraphics[scale=0.17]{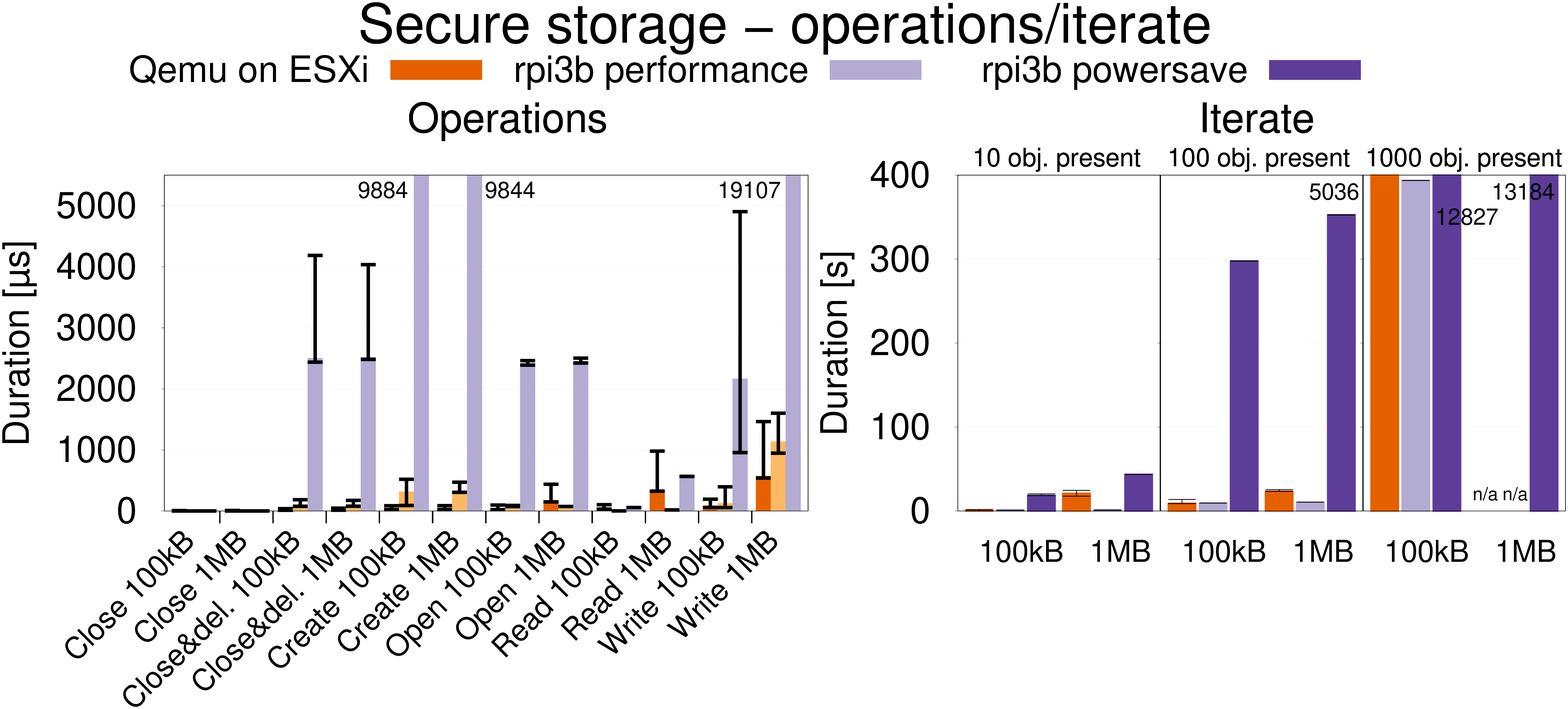}}
\caption{Secure storage: basic operations (left) and iteration (right)}\label{fig:performance_secure_storage_iter_storage}
\end{figure}

\textbf{Secure Storage: performance.}\label{subsec:secstorage} We evaluate the performance of \tz's secure storage via the corresponding GlobalPlatform's API implemented by \optee.
Specifically, we benchmark the cost of creating, writing, reading and closing objects inside the secure storage area, for two different object sizes (100KB and 1MB), although current memory allocator limitations prevented to cover some cases~\cite{allocmemissue1,allocmemissue2,allocmemissue3,allocmemissue4}.
Figure~\ref{fig:performance_secure_storage_iter_storage} (left) shows that closing and deleting objects are fast operations, and opening and writing are the slowest ones.
Iterating over objects in the secure storage (\emph{e.g.}, the execution of a \texttt{find} operation) is slow, up to a few hours in the worst case (Figure~\ref{fig:performance_secure_storage_iter_storage}, right).
Adding more objects in secure storage degrade the results even more (up to $2.01 \times object\_count\_ratio$).

\begin{figure}[t]
\centerline{\includegraphics[scale=0.17]{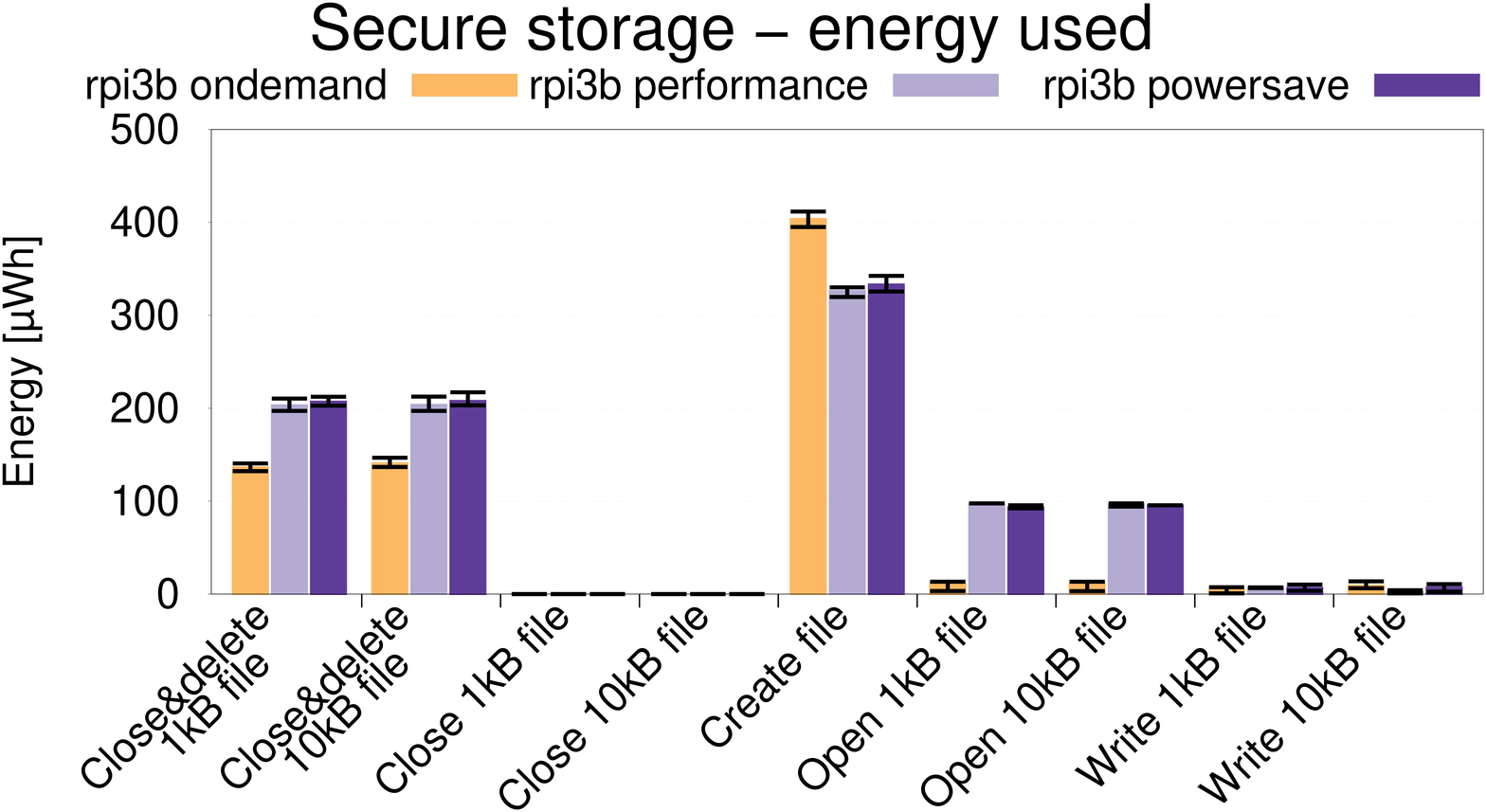}}
\caption{Secure storage: energy measurements for basic operations}\label{fig:power_secure_storage}
\end{figure}
\begin{figure}[t]
\centerline{\includegraphics[scale=0.17]{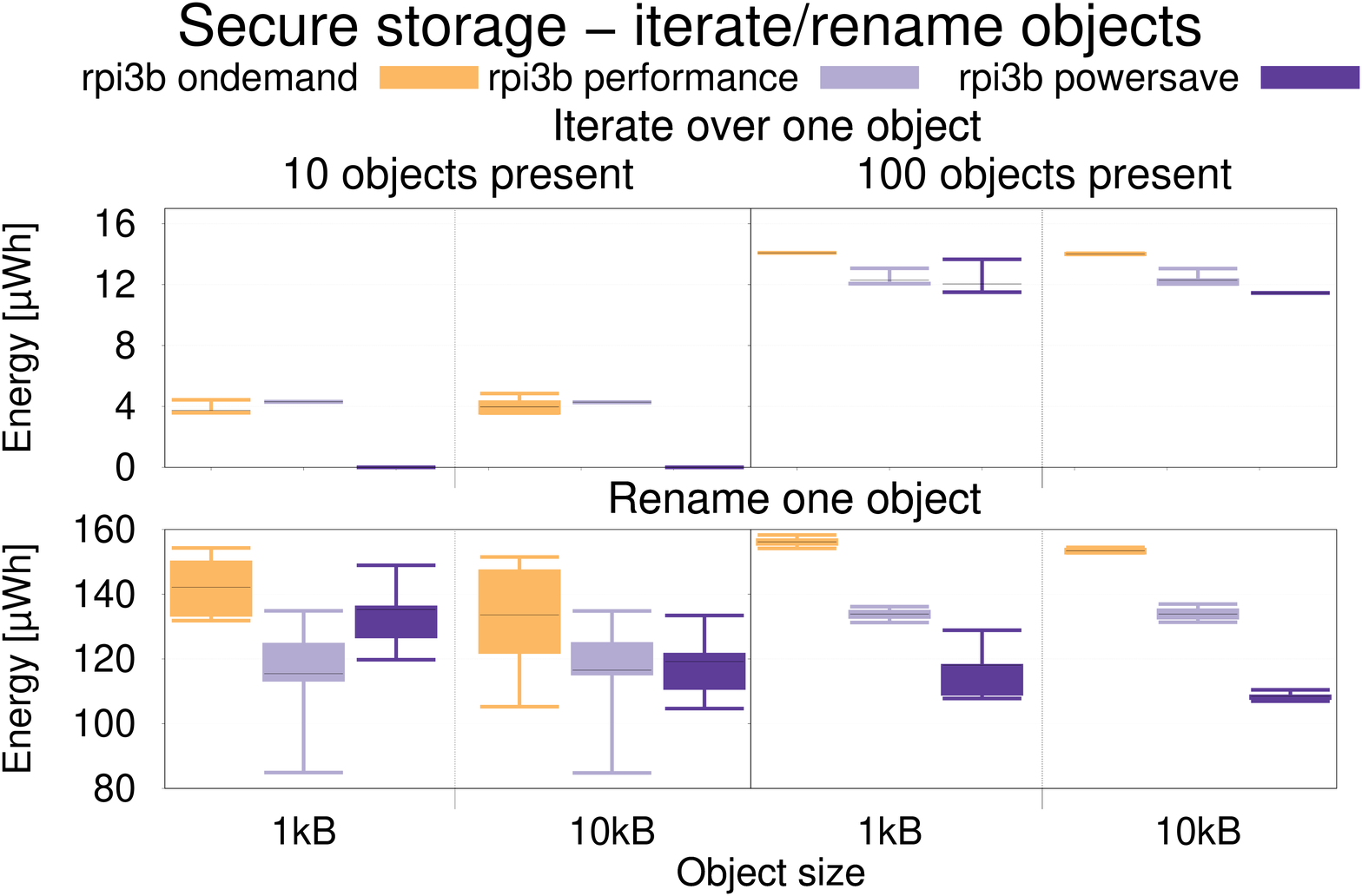}}
\caption{Secure storage, energy to iterate (top) and rename (bottom)}\label{fig:merged_power_secure_storage_iterate_rename}
\end{figure}

\begin{figure}[t]
\centerline{\includegraphics[scale=0.17]{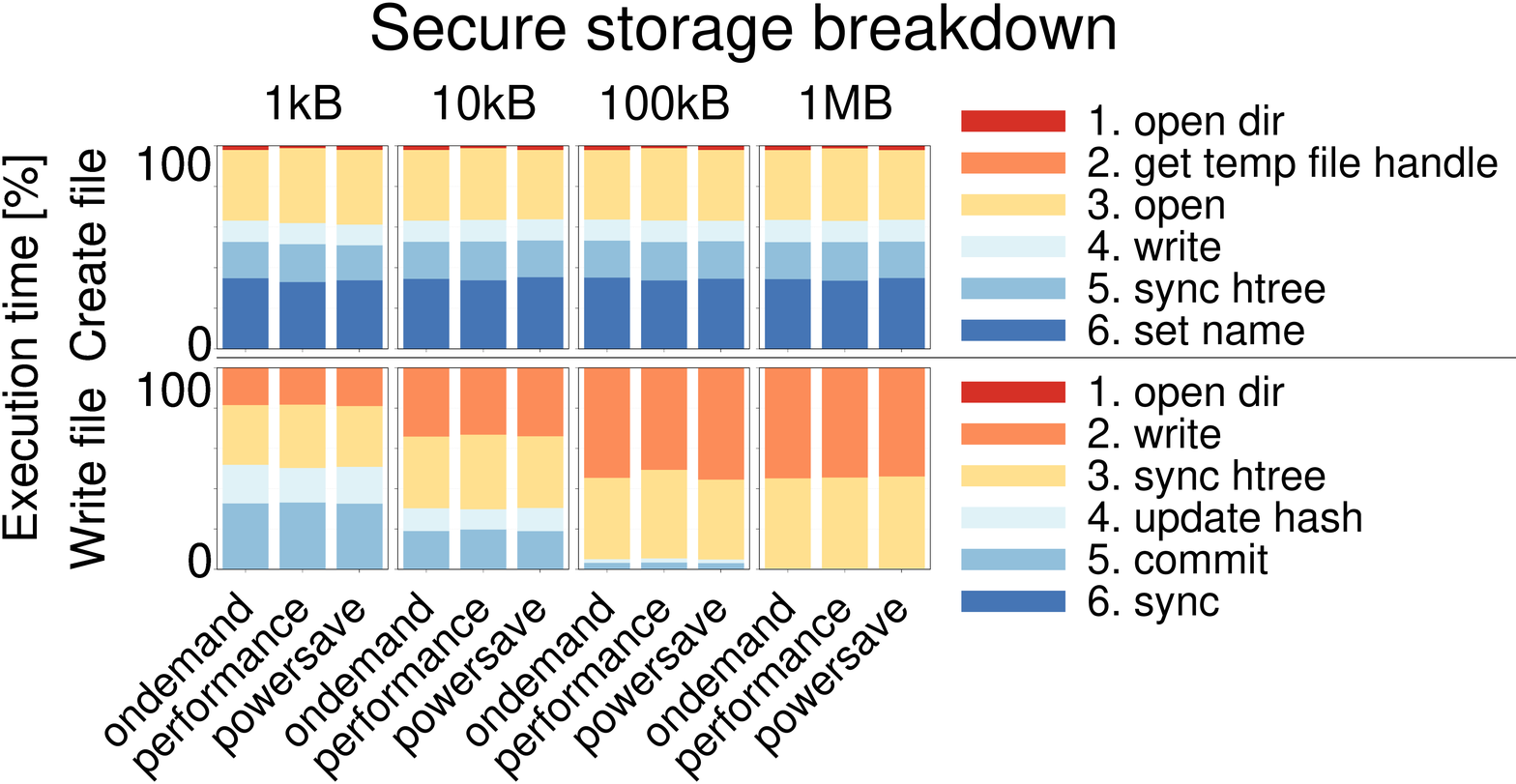}}
\caption{Secure storage breakdown for two operations: \textit{create} and \textit{write}}\label{fig:performance_secure_storage_micro}
\end{figure}	
\textbf{Secure storage: cost breakdown.} To understand how each low-level syscall affects the performance of a file-system inside the secure storage, we implemented a simple microbenchmark, inside \texttt{ree\_fs\_create} and \texttt{ree\_fs\_write}.
Specifically, these tests create and write data into a new object.
Figure~\ref{fig:performance_secure_storage_micro} shows a breakdown cost using stacked bars for writing and creating files.
These two functions are atomic and thus are surrounded by a monitor (mutex) which adds a considerable delay (not shown) regarding the \textit{write} operation. 
The impact is negligible on the \textit{create} operation.
We observe that opening the file and setting the filename accounts for the most time spent.

\textbf{Secure Storage: energy.} Being a feature often used by nomad devices with low energy autonomy, we deeply investigate its energy impacts.
Figure~\ref{fig:power_secure_storage} shows that creating objects is the most energy-demanding (up to 403µWh), irrelevant of the size. 
Power consumption of writing objects is dependent on their size.  
Interestingly, the \textit{ondemand} governor achieves slightly worse results when creating a file, whereas for closing and deleting files it stands out.
%
Figure~\ref{fig:merged_power_secure_storage_iterate_rename} shows the energy requirements to iterate over a single stored object (top)~\cite[Chapter 5.8]{teeinternalcorespecs} during enumeration of all stored objects in secure storage or rename (bottom) a single object, when additional 10 or 100 objects (of the same size) are already in the secure storage. 
We execute this test for 2 different file sizes (1kB and 10kB).
We observe that the energy required to iterate over a single object depends on the number of objects stored (in particular when using \texttt{performance} and \texttt{ondemand}), whereas the size of the object is irrelevant. 
	
%
	

\begin{figure}[t]
\centerline{\includegraphics[scale=0.16]{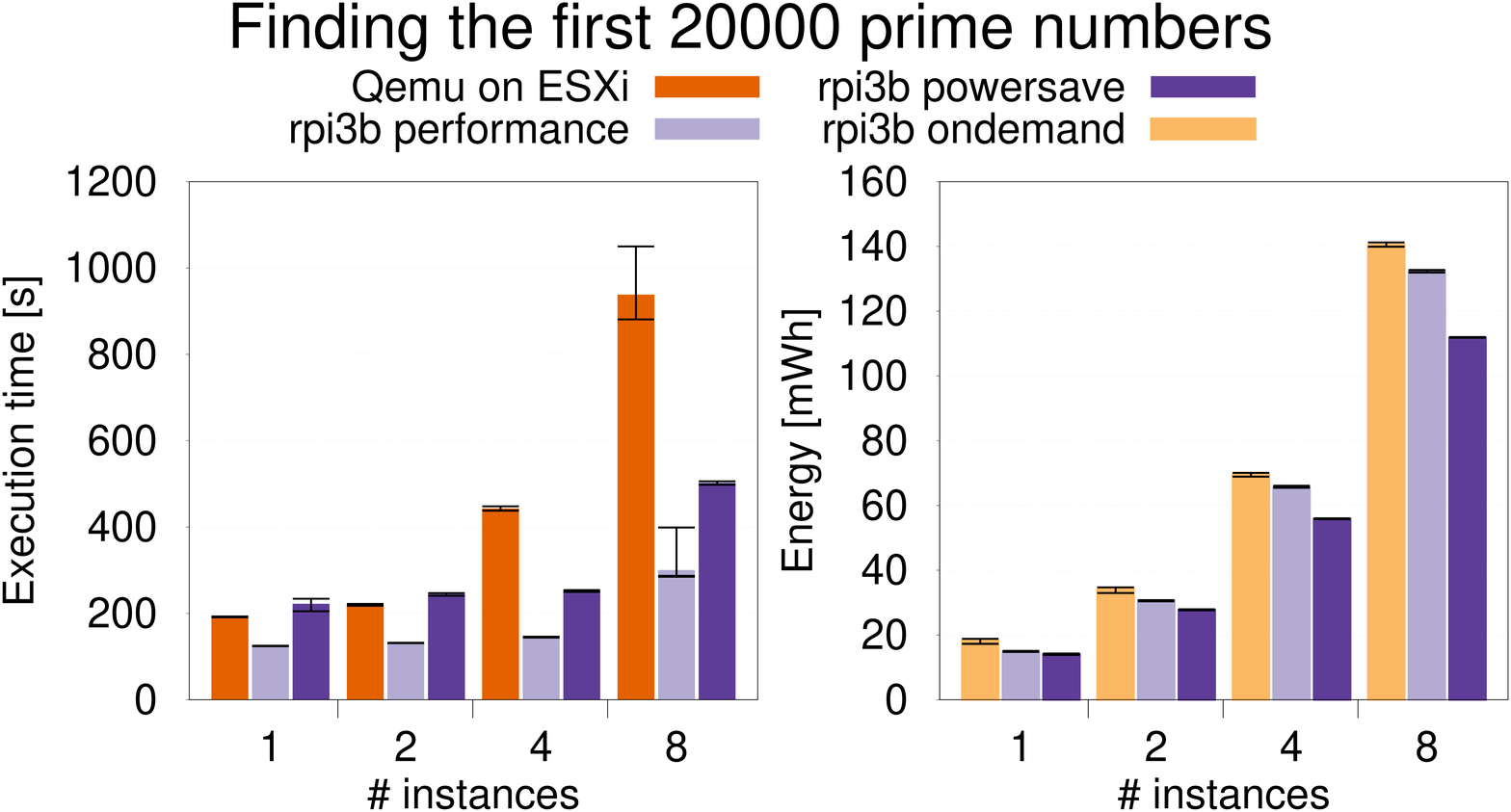}}
\caption{CPU benchmark: processing delay and energy requirements.}\label{fig:performance_power_cpu}
\end{figure}

\textbf{CPU Benchmarks.} To benchmark the raw performance of the \arm processors of our units, we implemented and deployed a single-threaded TA that executes a CPU-bound task, \emph{e.g.}, computes the first 20000 prime numbers. 
We run multiple instances concurrently, and while they execute we also gather energy measurements (for all cases minus the emulation mode).
Figure~\ref{fig:performance_power_cpu} presents these results.
As expected, the \texttt{performance} governor ensures the fastest computing time.
Due to emulation costs, the Qemu results are the worst ones.
As the number of instances exceed the available hardware cores, we observe an increase of energy consumption.
Overall, in this benchmark the \texttt{ondemand} governor is the most energy eager. 
This can be explained by the fact that adjusting the core frequencies (from 600MHz and 1.2GHz) seems to be a relatively costly operation~\cite{ibmworkloadsandgovernoreffects}.
	
\begin{figure}[t]
\centerline{\includegraphics[scale=0.17]{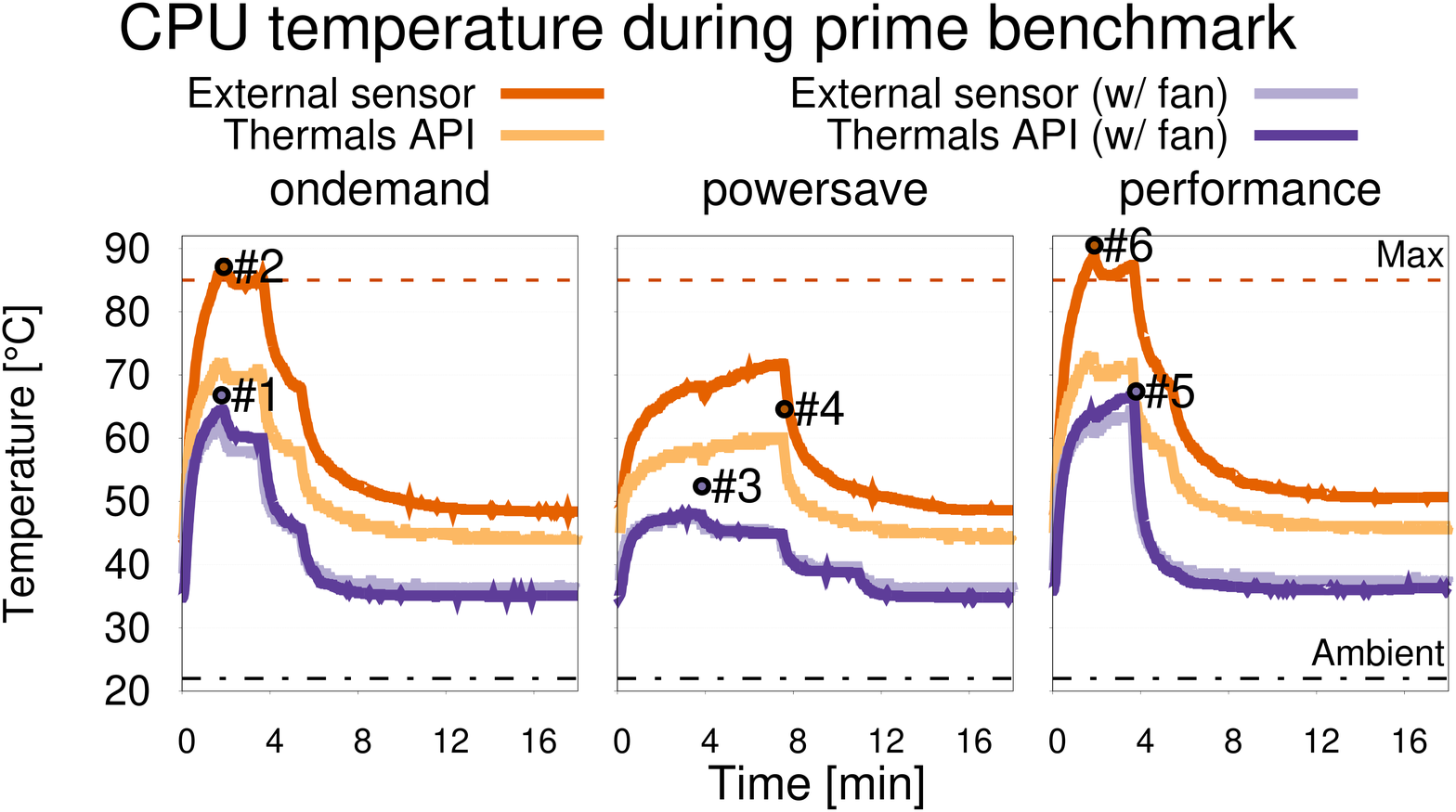}}
\caption{Evolution of CPU temperature with different cooling modes and governors.}\label{fig:cpu_temperature}
\end{figure}

\begin{figure}[t]
\centerline{\includegraphics[scale=0.20]{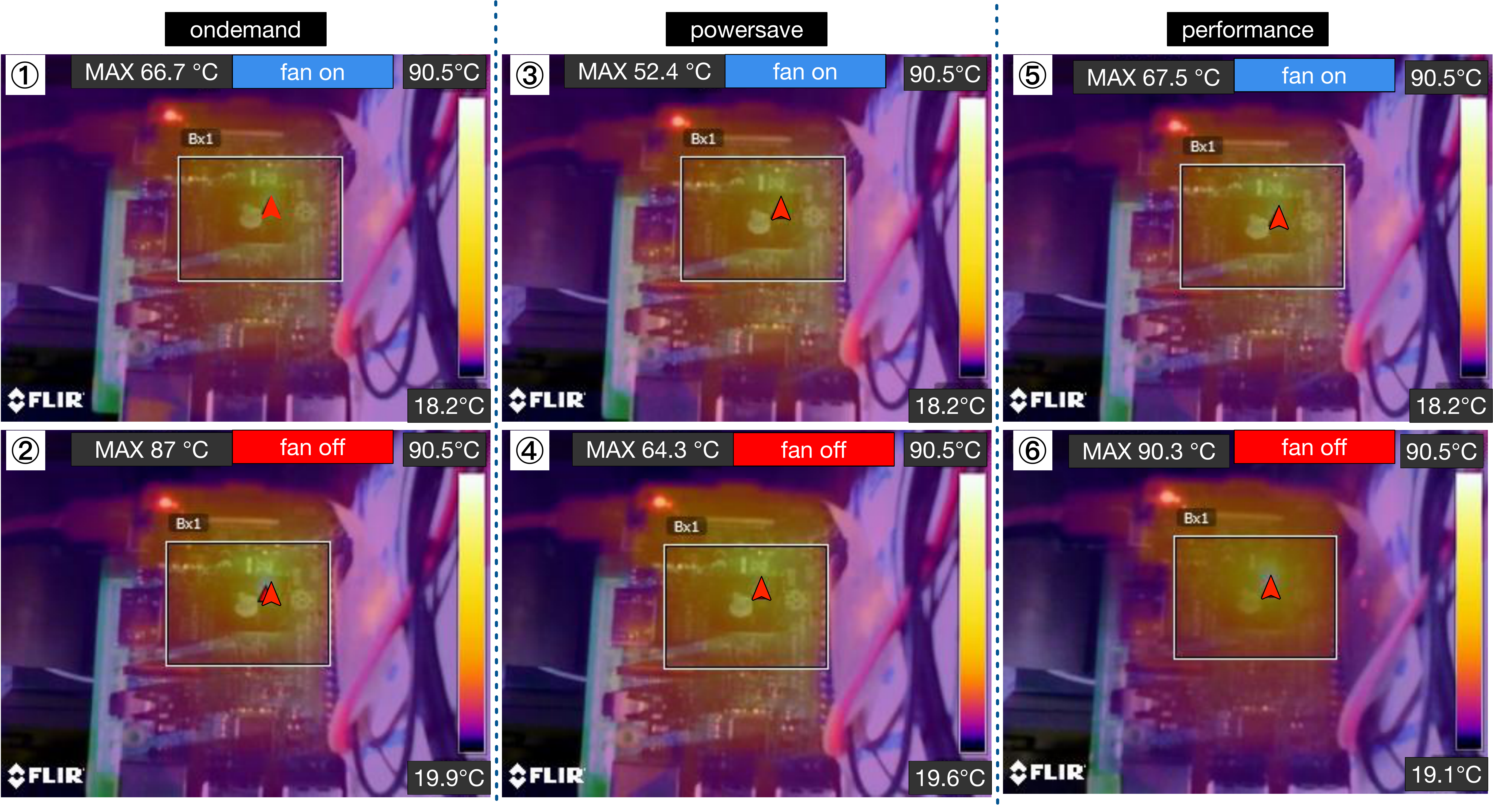}}
\caption{Raspberry Pi thermal behaviour during processor stress benchmarks.}\label{fig:thermal_pictures_cpu_temp}
\end{figure}

\textbf{Thermal benchmarks.} We conclude our evaluation by looking at the thermal envelope of the \emph{SoC}. 
To do so, we execute 8 concurrent instances of the prime benchmark inside \tz.
Figure~\ref{fig:cpu_temperature} presents the measurements fetched using the kernel's thermals API.
Additionally, we monitor the surface temperature of the chip using a Texas Instruments LM35 precision linear sensor with the help of an external micro controller.
Thermal conductivity between the \emph{SoC} and the LM35 is ensured by using a thermal compound (Arctic MX-4\cite{arcticmx4}).
The ambient temperature is of around 21.9°C.
Results returned by the LM35 are calibrated and checked at rest against a Fluke thermocouple, and against a Flir E4~\cite{flire4} thermal camera (see pictures in Figure~\ref{fig:thermal_pictures_cpu_temp}).
Marked points in Figure~\ref{fig:cpu_temperature} refer to measurements done using the thermal camera.
We observe a small margin of error of 3°C, and 
a discrepancy between the thermals API and the LM35 of over 15°C at times. 
This could be problematic because the measured surface temperature exceeds the rated continuous temperature of 85°C specified by the chip's manufacturer.
In this situation, the thermals API returns an incorrect temperature that is well below the acceptable temperature. As a consequence measures which should be taken to reduce the temperature, such as software thermal throttling, are not undertaken.
A passively cooled Raspberry Pi should therefore only operate in \texttt{powersave} mode or risk being hardware throttled or worse, suffer damage. 
An actively cooled system on the other hand can operate in any mode and stay well within acceptable conditions, even without additional heat sink.
Once the maximal temperature is reached, recovery time is around 8 minutes when passively cooled and less than a minute with active cooling.

\newpage
\section{Lessons Learned}
\label{sec:lessons}
This section reports on a few lessons learned during this experimental work.

\textbf{Memory limitations.}
By default, 32MB are dedicated to \optee, of which: 1MB for TEE memory, 1MB for PUB (non-secure RAM) memory, and the remaining 30MB for TAs.
Each TA has two compile-time options, \textit{TA\_STACK\_SIZE} and \textit{TA\_DATA\_SIZE} (in \textit{user\_ta\_header\_defines.h}), defining the stack size and heap size that can be utilized by a TA.
These values are set at very low values by default, 2kB and 32kB respectively~\cite{opteefaq}. 
For larger memory allocations, the TA's MMU L1 table must be set accordingly, as the default mapping is 1MB.
We were unable to allocate more than 3MB for a single TA, even with shared memory enabled. 
Consequently, the \optee  benchmark framework~\cite{opteebenchmarkframework} could not be used.

\textbf{Compliance to standards.}The GlobalPlatform's implementation in \optee is not error-free and some parts of the implementation do not comply fully with the specification.
For instance, the \textit{TEE\_BigIntAdd}~\cite[p. 252]{teeinternalcorespecs} function, contrary to its definition, does not allow to use the same pointers for both input and output~\cite{teebigintaddgitissue}.
Being relatively new, \optee  is improving rapidly. 
While this offers great advantages, such as mitigations against the latest attacks, it also introduces incompatibilities by deprecating older APIs. 
However, the GlobalPlatform consortium offers strong incentives for TEE vendors to comply with their API, which is unlikely to introduce breaking changes. 
Establishing this level of compliance ensures interoperability of TAs between existing TEE solutions which is undeniably of great interest to secure application developers.

\textbf{Developers toolchain.} The \optee framework groups all required dependencies in a single project while also including several components of its own, such as the secure kernel. 
This greatly facilitates development of secure application by reducing setup and development efforts. 
The \optee project includes a few TA examples and host applications, which are a good foundation to introduce the TEE paradigm.

\vspace{-10pt}
\section{Conclusion}\label{sec:conclusion}
\tz is a widely available technology that offers Trusted Execution Environment guarantees to low-energy devices.
The goal of this practical experience report was to uncover the performance of these systems.
To perform our experiments, we had to extend both secure and rich kernels so that secure timing measurements and thermal metrics could be fetched from within \tz, for which we provide detailed explanations in Appendix~\ref{appendix:extending-kernel}.
Our work highlights several advantages as well as limitation of the currently available software platforms, such as the  \optee framework chosen in our case, to implement and deploy TAs.
We would like to point out two major limitations.
(1) the lack of several basic features inside the REE kernel for security reasons, which materialize in the lack of basic syscalls (\emph{e.g.} \texttt{fopen}, \texttt{msgget}).
For this reason, it is paramount to reduce syscall dependencies when developing TAs. 
(2), the current limitations regarding memory allocation and addressing, which could negatively affect the facility to deploy more complex TAs inside \tz.
We hope this work will provide useful insights to \tz software developers.
\vspace{-10pt}

\section*{Acknowledgments} 
{\small
The research leading to these results has received funding from the European Union’s Horizon 2020 research and innovation programme under the LEGaTO Project (\url{legato-project.eu}), grant agreement No 780681.
}
\newpage
{\footnotesize
	\bibliographystyle{abbrv}
	\bibliography{main}
}
\newpage
\appendix
\section{Appendix: Extending the Kernel}\label{appendix:extending-kernel}

First, a new file containing the syscall used to retrieve the processor temperature \texttt{getcputemp} is created.


\lstset{language=C,caption={linux/custom/custom.c},captionpos=b}
\begin{lstlisting}
// populates temp with the CPU temperature in [m degC]
SYSCALL_DEFINE1(getcputemp, unsigned long *, temp)
{
   struct thermal_zone_device *tzd;
   // The name "bcm2835_thermal" is obtained
   // from /sys/class/thermal/thermal_zone0/type
   tzd = thermal_zone_get_zone_by_name("bcm2835_thermal");
   if (IS_ERR(tzd))
     return 1;
   thermal_zone_get_temp(tzd, &temp);
   return 0;
}
\end{lstlisting}

This file must be referenced in the main kernel Makefile:

\lstset{language=C,caption={linux/custom/Makefile},captionpos=b}
\begin{lstlisting}
core-y += kernel/ [...]  custom/
\end{lstlisting}

The syscall must be included in \texttt{syscalls.h}:

\lstset{language=C,caption={linux/include/linux/syscalls.h},captionpos=b}
\begin{lstlisting}
asmlinkage long sys_getcputemp(unsigned long *temp);
\end{lstlisting}

The \texttt{CALL} macro is used in \texttt{unistd.h}:

\lstset{language=C,caption={linux/arch/arm/kernel/calls.S},captionpos=b}
\begin{lstlisting}
CALL(sys_getcputemp)
\end{lstlisting}

Use the next available syscall identifier:

\lstset{language=C,caption={linux/arch/arm/include/uapi/asm/unistd.h},captionpos=b}
\begin{lstlisting}
#define __NR_getcputemp (__NR_SYSCALL_BASE+394)
\end{lstlisting}

In the following file and in addition to the modification listed above, note that \textit{\_\_NR\_syscalls} must be incremented by one.

\lstset{language=C,caption={linux/include/uapi/asm-generic/unistd.h},captionpos=b}
\begin{lstlisting}
#define __NR_getcputemp 288
__SYSCALL(__NR_getcputemp, sys_getcputemp)
\end{lstlisting}

At this point the new syscall is available to all user-mode applications running in TEE (Figure~\ref{fig:trustzone_components}-\ding{204} and Figure~\ref{fig:trustzone_components}-\ding{205}). 
This syscall is then exposed in the REE kernel, tee-supplicant and the TEE kernel as if it were an official GlobalPlatform's API function definition.

\lstset{language=C,caption={optee\_os/lib/libutee/include/tee\_api.h},captionpos=b}
\begin{lstlisting}
unsigned long TEE_GetCpuTemperature(void);
\end{lstlisting}

The \textit{\_\_NR\_syscalls} value must be modified to account for the new syscall:

\lstset{language=C,caption={linux/arch/arm/include/asm/unistd.h},captionpos=b}
\begin{lstlisting}
#define __NR_syscalls <INCREASE_BY_ONE>
\end{lstlisting}

The TEE function is a wrapper for the corresponding libutee implementation:

\lstset{language=C,caption={optee\_os/lib/libutee/tee\_api.c},captionpos=b}
\begin{lstlisting}
unsigned long TEE_GetCpuTemperature(void)
{
        unsigned long ret;

        TEE_Result res = utee_get_temperature(&ret);

        if (res != TEE_SUCCESS)
                TEE_Panic(res);

        return ret;
}
\end{lstlisting}

\textit{TEE\_SCN\_MAX} must also be increased accordingly and the call is given the next unique identifier (71 in our case):

\lstset{language=C,caption={optee\_os/lib/libutee/include/tee\_syscall\_numbers.h},captionpos=b}
\begin{lstlisting}
#define TEE_SCN_GET_TEMPERATURE 71
#define TEE_SCN_MAX <INCREASE_BY_ONE>
\end{lstlisting}

The utee syscall is declared in \texttt{utee\_syscalls.h} and linked to its unique identifier:

\lstset{language=C,caption={optee\_os/lib/libutee/include/utee\_syscalls.h},captionpos=b}
\begin{lstlisting}
TEE_Result utee_get_temperature(unsigned long *temp);
\end{lstlisting}

\lstset{language=C,caption={optee\_os/lib/libutee/arch/arm/utee\_syscalls\_asm.S},captionpos=b}
\begin{lstlisting}
UTEE_SYSCALL utee_get_temperature, TEE_SCN_GET_TEMPERATURE, 1
\end{lstlisting}

Add the syscall entry in \texttt{arch\_svc.c}. The trailing comma is required.

\lstset{language=C,caption={optee\_os/core/arch/arm/tee/arch\_svc.c},captionpos=b}
\begin{lstlisting}
SYSCALL_ENTRY(syscall_get_temperature),
\end{lstlisting}

\lstset{language=C,caption={optee\_os/core/include/tee/tee\_svc.h},captionpos=b}
\begin{lstlisting}
TEE_Result syscall_get_temperature(unsigned long *temp);
\end{lstlisting}

This function serves as a wrapper to the REE kernel syscall used to retrieve the temperature:

\lstset{language=C,caption={optee\_os/core/tee/tee\_svc.c},captionpos=b}
\begin{lstlisting}
#include <kernel/tee_temperature.h>
TEE_Result syscall_get_temperature(unsigned long *temp)
{
        tee_ta_get_temperature(temp);

        return TEE_SUCCESS;
}
\end{lstlisting}

A new file is created:

\lstset{language=C,caption={optee\_os/core/include/kernel/tee\_temperature.h},captionpos=b}
\begin{lstlisting}
#ifndef TEE_TEMPERATURE_H
#define TEE_TEMPERATURE_H

#include "tee_api_types.h"

TEE_Result tee_ta_get_temperature(unsigned long *temp);

#endif
\end{lstlisting}

This function is called in \circled{8} and triggers a REE world switch \circled{7}:

\lstset{language=C,caption={optee\_os/core/arch/arm/kernel/tee\_temperature.c},captionpos=b}
\begin{lstlisting}
#include <compiler.h>
#include <string.h>
#include <stdlib.h>
#include <optee_msg.h>
#include <kernel/thread.h>
#include <kernel/tee_temperature.h>

TEE_Result tee_ta_get_temperature(unsigned long *temp)
{
        TEE_Result res;
        struct optee_msg_param params;

        memset(&params, 0, sizeof(params));
        params.attr = OPTEE_MSG_ATTR_TYPE_VALUE_OUTPUT;
        res = thread_rpc_cmd(OPTEE_MSG_RPC_CMD_GET_TEMPERATURE, 1, &params);

        if (res == TEE_SUCCESS) {
                *temp = params.u.value.a;
        }

        return res;
}
\end{lstlisting}

The following line in added in \texttt{sub.mk}:

\lstset{language=C,caption={optee\_os/core/arch/arm/kernel/sub.mk},captionpos=b}
\begin{lstlisting}
srcs-y += tee_temperature.c
\end{lstlisting}

A new message used to retrieve the temperature via RPC is declared:

\lstset{language=C,caption={optee\_os/core/include/optee\_msg.h},captionpos=b}
\begin{lstlisting}
// [out] temperature
#define OPTEE_MSG_RPC_CMD_GET_TEMPERATURE 21
\end{lstlisting}

The same is done in another file:

\lstset{language=C,caption={linux/drivers/tee/optee/optee\_msg.h},captionpos=b}
\begin{lstlisting}
// [out] temperature
#define OPTEE_MSG_RPC_CMD_GET_TEMPERATURE 21
\end{lstlisting}

This function is declared inside the REE kernel:

\lstset{language=C,caption={linux/drivers/tee/optee/rpc.c},captionpos=b}
\begin{lstlisting}
#include <linux/syscalls.h>

static void handle_get_temperature(struct optee_msg_arg *arg)
{
        unsigned long cputemperature;

        // Linux kernel syscall
        if (sys_getcputemp(&cputemperature))	 {
        	arg-ret = TEEC_ERROR_GENERIC;
        	return;
        }

        arg->params[0].u.value.a = cputemperature;
        arg->ret = TEEC_SUCCESS;
        return;
}
\end{lstlisting}

In the same file, \texttt{handle\_rpc\_func\_cmd} is modified by adding a case to handle the new RPC request:

\lstset{language=C,caption={linux/drivers/tee/optee/rpc.c},captionpos=b}
\begin{lstlisting}
case OPTEE_MSG_RPC_CMD_GET_TEMPERATURE:
		handle_get_temperature(arg);
break;
\end{lstlisting}

After rebuilding the TEE client and kernel, the new syscall can be used as such from any TA \circled{9}:

\lstset{language=C,caption={Usage from TA},captionpos=b}
\begin{lstlisting}
float tempC = TEE_GetCpuTemperature() / 1000.0f;
\end{lstlisting}

This solution perfectly illustrates a workaround to the starvation of the REE world caused by the execution of the TEE.

In order to accomplish the secure storage micro benchmark, it was required to measure monotonic time, store and retrieve these measurements.
The common denominator between the TEE kernel and the host application is the REE kernel. For this reason, it was decided to store measurements in the REE kernel, from which they could be gathered by the host application. Three syscalls were added in the REE kernel and made available in the TEE kernel.

These are first declared:

\lstset{language=C,caption={linux/arch/arm/include/uapi/asm/unistd.h},captionpos=b}
\begin{lstlisting}
#define __NR_ktraceadd (__NR_SYSCALL_BASE+396)
#define __NR_ktraceget (__NR_SYSCALL_BASE+397)
#define __NR_ktracereset (__NR_SYSCALL_BASE+398)
\end{lstlisting}

\lstset{language=C,caption={linux/arch/arm/kernel/calls.S},captionpos=b}
\begin{lstlisting}
CALL(sys_ktraceadd)
CALL(sys_ktraceget)
CALL(sys_ktracereset)
\end{lstlisting}

\lstset{language=C,caption={linux/include/linux/syscalls.h},captionpos=b}
\begin{lstlisting}
// save the current time as the specified id
asmlinkage long sys_ktraceadd(unsigned long id); 
// returns the id+sec+ns of the requested index
asmlinkage long sys_ktraceget(unsigned long index, unsigned long* id,
                                   unsigned long* sec, unsigned long* ns); 
asmlinkage long sys_ktracereset(void);
\end{lstlisting}

Implementation is stored in a separate file:

\lstset{language=C,caption={linux/custom/custom.c},captionpos=b}
\begin{lstlisting}
#include <linux/kernel.h>
#include <linux/syscalls.h>
#include <linux/timekeeping.h>
#include <linux/slab.h>

#define MAX_KTRACE_ENTRIES 30
unsigned char ktrace_entries = 0;

struct ktraceadd_e {
	unsigned long id;
	struct timespec64 ts;
};

struct ktraceadd_e* ktraceadd_d;

SYSCALL_DEFINE1(ktraceadd, unsigned long, id)
{
	struct timespec64 ts;
	ts = ns_to_timespec64(ktime_get_ns());
	
	if (!ktraceadd_d) {
		ktraceadd_d = kmalloc(sizeof(struct ktraceadd_e)*MAX_KTRACE_ENTRIES,
		                         GFP_KERNEL | GFP_NOWAIT);
	}
	
	if (ktrace_entries < MAX_KTRACE_ENTRIES) {
		memcpy((void*)&ktraceadd_d[ktrace_entries].id, (void*)&id, sizeof(unsigned long));
		memcpy((void*)&ktraceadd_d[ktrace_entries].ts, (void*)&ts, sizeof(struct timespec64));
		ktrace_entries++;
		return 0;
	}

	return 1;
}

SYSCALL_DEFINE4(ktraceget, unsigned long, index, unsigned long*, id, unsigned long*, sec, unsigned long*, ns)
{
	if (ktraceadd_d && index >= 0 && index < ktrace_entries) {
		*id = ktraceadd_d[index].id;
		*sec = ktraceadd_d[index].ts.tv_sec;
		*ns = ktraceadd_d[index].ts.tv_nsec;
		return 0;
	}
	
	return 1;
}

SYSCALL_DEFINE0(ktracereset)
{
	ktrace_entries = 0;
	return 0;
}
\end{lstlisting}

Next, three available syscalls identifiers are used.
In the same file, \textit{\_\_NR\_syscalls} must be incremented by three.

\lstset{language=C,caption={linux/include/uapi/asm-generic/unistd.h},captionpos=b}
\begin{lstlisting}
#define __NR_ktraceadd 290
__SYSCALL(__NR_ktraceadd, sys_ktraceadd)
#define __NR_ktraceget 291
__SYSCALL(__NR_ktraceget, sys_ktraceget)
#define __NR_ktracereset 292
__SYSCALL(__NR_ktracereset, sys_ktracereset)
\end{lstlisting}

The \textit{\_\_NR\_syscalls} value must be modified to account for the new syscalls:

\lstset{language=C,caption={linux/arch/arm/include/asm/unistd.h},captionpos=b}
\begin{lstlisting}
#define __NR_syscalls <INCREASE_BY_THREE>
\end{lstlisting}

These functions can now be invoked from any REE user-mode application.
Instrumentation tests \textit{test1} and \textit{test2} are added directly from the host application using \circled{3}, and then retrieved and displayed.

\lstset{language=C,caption={Host application usage example},captionpos=b}
\begin{lstlisting}
#include <unistd.h>
#include <sys/syscall.h>
#include <time.h>

// As defined in
// optee/linux/include/uapi/asm-generic/unistd.h
#define SYSCALL_KTRCEADD 290
#define SYSCALL_KTRCEGET 291
#define SYSCALL_KTRCERESET 292

printf("Calling SYSCALL_KTRCERESET\n");
syscall(SYSCALL_KTRCERESET);

printf("Calling SYSCALL_KTRCEADD 0\n");
syscall(SYSCALL_KTRCEADD, "test1");
sleep(1);
printf("Calling SYSCALL_KTRCEADD 1\n");
syscall(SYSCALL_KTRCEADD, "test2");

char kget_name[20];
unsigned long kget_sec;
unsigned long kget_ns;

for (int i = 0; i  < 2; ++i) {
	syscall(SYSCALL_KTRCEGET, i, &kget_name, &kget_sec, &kget_ns);
	printf("SYSCALL_KTRCEGET index %d: name=%s sec=%ld ns=%ld\n",
	        i, kget_name, kget_sec, kget_ns);
}
\end{lstlisting}

Trace calls can then be added anywhere in the REE core \circled{8}. Once called, an RPC is made \circled{7} to the REE kernel.
For example:

\lstset{language=C,caption={optee\_os/core/tee/tee\_ree\_fs.c},captionpos=b}
\begin{lstlisting}
#include <kernel/tee_ktrace.h>

static TEE_Result ree_fs_create([...])
{
	TEE_Result res;
	
	// Measuring instrumentation delay
	tee_ta_add_ktrace(200);
	tee_ta_add_ktrace(201);

	// Measuring time taken to enter the monitor
	mutex_lock(&ree_fs_mutex);
	tee_ta_add_ktrace(202);
	
	// [...]
}
\end{lstlisting}

\end{document}